\documentclass{article}
\usepackage{ulem}  
\usepackage{epsfig,amsmath,amsfonts,amssymb,graphicx}

\newcommand{\fivesp}{\ \ \ \ \ }

\begin{document}
\title{Chiral Dynamics in Photo-Pion Physics: Theory, Experiment, and Future
Studies at the HI$\gamma$S Facility}
\markboth{Aron M. Bernstein \textit{et al.}}{Chiral Dynamics in Photo-Pion Physics: Theory, Experiment, and Future
Studies at the HI$\gamma$S Facility}
\author{Aron M. Bernstein
\\Department of Physics and Laboratory for Nuclear Science,\\ Massachusetts Institute of Technology, \\Cambridge, MA 02139, USA \and Mohammad W. Ahmed \\Department of Physics and \\Triangle Universities Nuclear Laboratory, \\Duke University, Durham, NC, 27708-0308, USA \and Sean Stave \\Department of Physics and \\Triangle Universities Nuclear Laboratory, \\Duke University, Durham, NC, 27708-0308, USA \and Ying K. Wu \\Department of Physics and \\Triangle Universities Nuclear Laboratory, \\Duke University, Durham, NC, 27708-0308, USA \and Henry R. Weller \\Department of Physics and \\Triangle Universities Nuclear Laboratory, \\Duke University, Durham, NC, 27708-0308, USA}


\maketitle
\begin{abstract}
A review of photo-pion experiments on the nucleon in the near threshold region is presented. Comparisons of the results are made with the predictions of the low energy theorems of QCD calculated using chiral perturbation theory (ChPT) which is based on the  spontaneous breaking of chiral symmetry as well as its explicit breaking due to the finite quark masses.
As a result of the vanishing of the threshold amplitudes in the chiral limit, the experiments are difficult since the cross sections are small. Nevertheless the field has been brought to a mature stage of accuracy and sensitivity. The accomplishments and limitations of past experiments are discussed.  Future planned experiments at Mainz and HI$\gamma$S using polarization observables are discussed  as a more rigorous test of theoretical calculations. Emphasis is given to the technical developments that are required for the HI$\gamma$S facility. It is shown that future experiments will provide more accurate tests of ChPT and will be sensitive to isospin breaking dynamics due to the mass difference of the up and down quarks.
\end{abstract}

\section{Introduction}\label{sec:intro}

The fundamental nature of threshold photo- and electro-pion production
processes on the nucleon (N) has long been recognized~\cite{Amaldi} due to the fact that as a result of the spontaneously hidden (broken) chiral symmetry in QCD~\cite{Nambu,Nambu-2,Nambu-3} the
pion is a Nambu-Goldstone Boson~\cite{DSM}. Early work with
the low energy theorems of current algebra made predictions for the s-wave
amplitudes for the $N(\gamma,\pi)$ reaction~\cite{LET,LET-2,LET-3,LET-4,KR}. Later, an effective-field-theory for QCD called 
chiral perturbation theory (ChPT) was
employed~\cite{DSM,ChPT,ChPT-2,ChPT-3,ChPT-4,workshop,ChPT-review, ChPT-Baryons}. This is an expansion in momentum and pion (light quark) masses which
are small compared to the chiral symmetry breaking scale of $\simeq$ 1 GeV.
Since the perturbation series is truncated at a finite order the effect of
the higher order  neglected terms can be estimated to determine the
theoretical error. The main point is  that these calculations should be a
close  approximation to QCD, and that any discrepancy between ChPT theory
and experiment which are larger then the estimated errors must be taken
seriously.

ChPT with one-pion-loop corrections (``one-loop order'') has been used to
calculate the amplitudes for threshold photo- and electro-pion
production from the nucleon~\cite{loop,loop-2,loop-3,loop2,BKGM,BKMZ}. This theoretical development was driven by a series
of threshold photoproduction experiments with high-duty-cycle accelerators
carried out at Mainz~\cite{Mainz,Mainz-2,Schmidt} and Saskatoon~\cite{Sask}. 
Today there is impressive overall agreement between the ChPT calculations
and the data. This agreement with a large body of data supports the concept of spontaneous chiral symmetry hiding in QCD~\cite{DSM,ChPT,ChPT-2,ChPT-3,ChPT-4,workshop,ChPT-review, ChPT-Baryons,AB_workshop}.

The developments in photo-pion production followed similar work in
the $\pi$N sector which is closely linked to photo-pion production by unitarity, or to put
it more physically, by $\pi$-N interactions in the final states. As will be discussed, this leads to 
exciting opportunities to measure $\pi$-N phase shifts in properly designed photo-pion experiments.
This requires transversely polarized targets to be fully exploited, and is one of the major issues which we expect to be experimentally studied for the first time in the next few years.
Calculations of $\pi$-N interactions started with the current algebra predictions for 
pion-hadron scattering lengths~\cite{W1}. Later calculations were performed with
effective chiral Lagrangians~\cite{W2} followed by a series of ChPT
calculations~\cite{ChPT-Baryons,FM,FM-2} in low-energy $\pi$N
scattering~\cite{ChPT-Baryons,FM,FM-2} which are in good agreement with experiment.
Of particular importance is the precise measurement of $\pi$-N scattering lengths in pionic hydrogen and deuterium~\cite{Gotta}.
These have been successfully interpreted with ChPT and require isospin breaking due to the Coulomb interaction~\cite{Meissner-pi-D,Meissner-pi-D-2,Meissner-pi-D-3}.

The modern view is that  QCD exhibits chiral symmetry when the light
quark masses are set to zero (the chiral limit) in the Lagrangian. The
absence of mass-degenerate parity doublets in the spectra of hadrons suggests
that this symmetry is spontaneously broken (or hidden). The symmetry is not
lost but appears in the form of massless, pseudoscalar, Nambu-Goldstone Bosons.
Spontaneous symmetry breaking is well known in condensed matter physics,
e.g. magnetic domains in iron which break the rotational symmetry of the
Coulomb interaction. In this case, the Goldstone Bosons are spin waves or
magnons. In QCD, the small non-zero light-quark masses explicitly break the
chiral symmetry of the Lagrangian, with the result that the pion, eta and
kaon have non-zero masses. Nevertheless, these eight pseudoscalar mesons are the 
lightest hadrons coming into the mass gap $\lesssim$ 1 GeV.

As the lightest hadron, the pion best approximates the ideal Nambu-Goldstone
Boson. In the chiral limit, where the light quark and pion masses $\rightarrow$ 0, the 
pion would not interact with hadrons at  low energies
(i.e. the s-wave scattering lengths would vanish). In reality, the small
mass makes the low energy interaction weaker than a typical strong interaction, but non-zero. The
near-threshold interactions are important to measure since they are an
explicit effect of chiral symmetry breaking and have been calculated by
ChPT. In the future, lattice gauge theory is expected to also make accurate predictions.
Since the occurrence of the quasi Nambu-Goldstone Bosons signifies spontaneous
chiral symmetry breaking in QCD, their low-energy interactions with other hadrons, their electromagnetic production and decay amplitudes as well as
their internal properties (e.g. radii, polarizabilities, decay) will serve
as fundamental probes of the chiral structure of matter. These measurements
represent timely  issues since any disagreement between theory
and experiment represents a possible failure of QCD. These measurements
are often technical challenges for experimental physics. 

The most precise tests of chiral dynamics are in the mesonic sector involving the Nambu-Goldstone Bosons($\pi,\eta$,K)-particularly the pion, which is the lightest member of this family. In the past five years, there has been rapid progress in making such measurements. The recent NA48 high statistics experiment at CERN on the $K^{\pm} \rightarrow 
\pi^{+} \pi^{-} e^{\pm} \nu (K_{e4})$ and of the unitary cusp in the $ K^{\pm} \rightarrow \pi^{\pm} \pi^{0} \pi^{0}$ decays is the process of 
accurately measuring the s-wave  $\pi \pi$ scattering lengths~\cite{NA48}. These are  found  to be in agreement with ChPT calculations to two-loop order and with lattice calculations~\cite{pi-pi-theory}, and represent a critical test of chiral dynamics and of the basic assumption of symmetry which is spontaneously hidden and explicitly broken by the small, but finite, light quark masses. This agreement was achieved by also including the isospin breaking~\cite{pi-pi-theory}

In addition, experiments on pionic hydrogen and 
deuterium at PSI have measured the
 s-wave $\pi N$ scattering length~\cite{Gotta} which was found to be in
agreement with calculations provided that the Coulomb contribution to isospin breaking was taken into account~\cite{Meissner-pi-D,Meissner-pi-D-2,Meissner-pi-D-3}. On an equal chiral footing, the amplitude for neutral pion photoproduction
vanishes at low energies in the chiral limit. Existing data for this small
magnitude~\cite{Schmidt,Sask} are in reasonable agreement with ChPT calculations~\cite{loop,loop-2,loop-3, loop2}. Overall, these pion scattering and
production experiments strongly support our underlying concept of the pion as a quasi Nambu-Goldstone Boson, 
reflecting spontaneous chiral symmetry breaking in QCD~\cite{AB_workshop}.

Despite these successes, not all of the chiral predictions have been
properly tested. The long-standing prediction of Weinberg~\cite{W2} that the mass
difference of the up and down quarks leads to isospin breaking in $\pi N$
scattering (in addition to the electromagnetic contribution) is of special interest in chiral dynamics. The accuracies of the
completed experiments and of the model extractions from the deuterium pionic
atom do not yet permit a rigorous test of this fundamental prediction. An
interesting possibility is the use of the pion-photoproduction reaction
with polarized targets to measure the isospin-breaking predictions of low
energy $\pi^0 N$ scattering, which is related to the isospin-breaking
quantity $\frac{m_d -m_u}{m_d+m_u}\simeq 30\%$~\cite{W2,FM,FM-2,MSS}. This large value presents an
unusual experimental opportunity since the general order of magnitude of
the predicted isospin breaking is $\frac{m_d - m_u}{\Lambda_{QCD}}\simeq
2\%$. There have been claims of observing isospin breaking effects 
in medium-energy
$\pi N$ scattering experiments at several times this
magnitude~\cite{Gibbs,Gibbs-2}. These claims need independent testing! A proposed method to do this with photo-pion production~\cite{AB_lq} will be discussed in Sec.~\ref{sec:future}.

It is possible to characterize the successes of low energy
$\pi$N and photo- and electro-pion production experiments as verifying
that the pion is the approximate Nambu-Goldstone Boson of QCD, and that its 
low-energy production and interactions vanish in the chiral limit $m_{u} +
m_{d} \rightarrow 0$~\cite{AB_workshop}. We can  characterize the new generation of fully polarized  photo-pion experiments 
that will be carried out at the HI$\gamma$S facility and in Mainz as accurately
testing the concept of spontaneous chiral symmetry breaking in QCD
and the ChPT predictions based on this, and studying the consequences of $m_{d} -m_{u} > 0$. 

\section{Photopion Production Physics and $\pi$N Scattering\label{sec:photopion}}

\subsection{The Fermi-Watson (Final State Interaction) Theorem and Isospin Breaking Corrections }\label{sec:piN}
There is a deep connection between pion photoproduction and $\pi$N scattering. Formally, this occurs as a 
result of the unitarity of the S matrix. Physically, the connection is due to final-state interactions. In 
fact, since the photo-nucleon interaction is of small order $(e^{2}\simeq \alpha)$, the phase of the 
pion-photoproduction multipole amplitudes is defined by the
relationship:\footnote{The notation is that A stands for the electric (E) or magnetic (M) multipole. The final  state of the $\pi$N system in total angular momentum $j = l \pm 1/2$ (e.g, E$_{0+}$ means electric dipole with $l=0$ and $j=1/2$), and  isospin $I = 1/2$ or $3/2$. The total center-of-mass (CM) energy of the system $W$ is implicit.}
\begin{equation}
\label{FW}
 A_{l,j}^{2I}  = e^{i\delta_{l,j}^{2I} }  \widehat{A}_{l,j}^{2I}
\end{equation}
where  $\widehat{A}_{l,j}^{2I}$ are real functions of the CM  energy,
and can be identified as the multipole matrix elements 
for $\gamma p \rightarrow \pi N(I) $ in the absence of final state 
$\pi N$ interactions. Equation~\ref{FW} is the Fermi-Watson, or final state interaction, theorem when $\delta$ is identified as the elastic scattering
 $\pi$N phase shift~\cite{FW,FW-2}. In 
comparison to the elastic $\pi$N scattering S matrix = $e^{2i\delta_{l,j}^{2I}}$, Eq.~\ref{FW} shows that the phases of the $\pi N$ multipoles enter with half the magnitude that they do in $\pi N$ scattering.  
The Fermi-Watson theorem is very general, based on time-reversal invariance, three channel unitarity and isospin 
conservation. It is only valid below the two-pion threshold since the assumption of three channel unitarity is violated at higher energies. Strictly speaking it is not a true theorem since  it must be modified if isospin conservation is not strictly valid.

When the Fermi-Watson theorem was derived, 
quarks were not known. It was assumed that isospin violation was caused only by electromagnetic effects 
of order $e^{2}= \alpha$ and could be neglected. However, we now know there is an additional 
isospin-breaking mechanism due to the mass difference of the up and down quarks \cite{FM,FM-2,W2}. The modification 
of the \nobreak{Fermi-Watson} theorem due to this mechanism has been worked out \cite{AB_workshop,AB_lq,Anant}. To generalize to 
the isospin breaking case, one can write the S-matrix for each quantum state $ j = l \pm 1/2, I$ and total CM energy $W$ as:
\begin{equation}
\label{S_IS}
S_{l,j}^{2I}(\gamma p \rightarrow \pi N) =\left(\begin{array}{lcr}
1& i A_{l,j}^{1}& iA_{l,j}^{3}\\
   &\cos\psi \, e^{2i\delta_{l,j}^{1}}& i \sin\psi\, 
e^{i(\delta_{l,j}^{1}+\delta_{l,j}^{3})}\\
& & \cos\psi \, e^{2i\delta_{l,j}^{3}}
\end{array}\right)
\end{equation}
where sin $\psi$  represents an isospin-violating term and $ \psi$ is a 
real number which is a function $W$. For $\psi \rightarrow 0$, the isospin violation vanishes. 
The form of the 3x3 and 2x2 $\pi N$  portions (lower right) of the S 
matrix have been chosen to be separately unitary and time-reversal 
invariant. Applying the unitary constraint $S^{\dagger}S = SS^{\dagger} = 1$,  and 
assuming the weakness of the electromagnetic interaction by dropping  
terms of order e$^2$, one obtains \cite{AB_workshop,AB_lq}:
\begin{equation}
\begin{array}{rl}
\label{eq:M_IS}
 A_{l,j}^{1}(\psi)& = e^{i\delta_{l,j}^{1}}  [A_{l,j}^{1}(\psi=0)
 \cos\frac{\psi}{2} 
 + i A_{l,j}^{3}(\psi=0)  \sin \frac{\psi}{2} ] \\
A_{l,j}^{3}(\psi)   & =  e^{i\delta_{l,j}^{3} } [A_{l,j}^{3}(\psi=0) \cos
\frac{\psi}{2}+ i A_{l,j}^{1}(\psi=0) \sin \frac{\psi}{2} ]
\end{array}
\end{equation}
where $ A_{l,j}^{2I(\psi)}(A_{l,j}^{2I}(\psi=0))$ are the the $\gamma p \rightarrow \pi N $  multipoles   with (without) isospin breaking interactions. As $\psi \rightarrow 0$, 
the isospin violation vanishes and the Fermi-Watson theorem is recovered. 
Strictly speaking the isospin label I for 
$ A_{l,j}^{2I}(\psi)$ should not be included since it is not conserved. However, since the isospin violations are relatively small in magnitude, this approximate label is appropriate. 
As a consequence of Eq. ~\ref{eq:M_IS}, 
the phases of the $\pi N$ multipoles should be measured, not 
calculated from the $\pi N$ phase shifts (using the Fermi-Watson theorem), 
as they are now. To our knowledge, only one such measurement of the phases 
of the photoproduction multipoles has been performed \cite{Grushin}, and 
that did not have the required precision to demonstrate  a violation of the
unmodified Fermi-Watson theorem.

\subsection{ Photopion Production in the Threshold Region and Isospin Violation}\label{sec:IS_thresh}

The effects of isospin breaking are dramatic
in the threshold region. First, there is a  significant 
charge-dependent energy difference between the thresholds for the
 $\gamma p \rightarrow \pi^{0} p~(\pi^{+}n)$ 
reactions (144.7 and 151.4 MeV). For the region below the $\pi^{+}n$ threshold, isospin is completely 
broken, since only one channel is open. However, even in this sub-threshold region, there is a strong 
influence of the charged-pion channel through the two-step $\gamma p \rightarrow \pi^{+}n \rightarrow \pi^{0}p$ 
reaction. Since the ratio of the electric dipole amplitudes for the neutral and charged pion channels $ E_{0+}(\gamma p \rightarrow  \pi^{+}n) /E_{0+}(\gamma p 
\rightarrow \pi^{0}p)\simeq -20$, the two step reaction is as 
strong as the direct  path. This leads to a unitary cusp in the $\gamma p 
\rightarrow \pi^{0}p$ reaction.  

The simplest dynamical way to understand the occurrence of the unitary cusp in the 
$\gamma^{*} p\rightarrow \pi^{0}p$ reaction (where $\gamma^{*}$ is a real or virtual photon) is to use the 
3-channel S matrix for the open channels $\gamma^{*} p, \pi^{0}p$ 
and $\pi^{+}n$ \cite{AB_lq} (similar to Eq. \ref{S_IS}). Applying the constraints of unitarity
and time-reversal invariance, one is then led to the coupled-channel result  
for the s-wave amplitude $E_{0+}(\gamma p \rightarrow \pi^{0}p)$: 
\begin{eqnarray}\label{eq:unitary_cusp}
E_{0+}(\gamma p \rightarrow \pi^{0}p:k) = e^{i\delta_{0}  }  [A(k) + 
i \beta q_{+} ] \nonumber \\
\mbox{ with }\beta= E_{0+}( \gamma p \rightarrow \pi^{+}n) \cdot 
a_{cex}(\pi^{+}n\rightarrow\pi^{0}p)
\end{eqnarray}
where $\delta_{0}$ is the s-wave $\pi^{0}p$  phase shift (predicted to 
be small), $A(k)$ is a smooth function of the photon energy $k$,  $\beta$
is the cusp  parameter, $a_{cex}(\pi^{+}n\rightarrow\pi^{0}p)$ 
is the charge exchange s-wave scattering length for the $\pi^{+}n \rightarrow \pi^{0}p$ reaction and 
$q_{+}$ is the $\pi^{+}$  CM momentum\footnote{With $a_{cex}(\pi^{+}n \rightarrow \pi^{0}p)$ in units of $1/m_{\pi^{+}}$ and $q_{+}$ in $m_{\pi^{+}}$ units.}
, which is continued below the
$\pi^{+}n$ threshold as $i\mid q_{+}\mid$. The cusp function 
$\beta q_{+}$ contributes to the real (imaginary) part of $E_{0+}$  
below (above) the $\pi^{+}n$ threshold. It is interesting that the constraints of unitarity show that 
the two-step $\gamma p \rightarrow \pi^{+}n\rightarrow 
\pi^{0}p$ mechanism is the most important energy dependence in  
the near-threshold $\gamma p \rightarrow \pi^{0}p$   reaction. This 
is in agreement with the predictions of ChPT \cite{loop,loop-2,loop-3,loop2}. 

The expected value of $\beta$ can be 
calculated \cite{AB_lq} on the basis of unitarity using Eq.~\ref{eq:unitary_cusp}.
Note that the sign of $\beta$ is observable, not just it's magnitude, and
agrees with what is expected (see below).
The best experimental
value of $a(\pi^{-}p\rightarrow\pi^{0}n) = -(0.122\pm 0.002) 
/m_{\pi}$, obtained from the observed width in the 1s state of pionic hydrogen
\cite{Gotta}, was used. This is in excellent agreement with ChPT
predictions of $-(0.130\pm0.006) /m_{\pi}$ \cite{s-pred}. Assuming
isospin is conserved, $a(\pi^{+}n \leftrightarrow \pi^{0}p) = -a(\pi^{-}p
\leftrightarrow \pi^{0}n)$. The latest measurement 
for  $E_{0+}({\gamma}p \rightarrow \pi^{+}n)= 28.06 \pm 0.27 \pm 0.45$\footnote{The units 
for $E_{0+}$ and $\beta$ are $10^{-3}/m_{\pi}$,  and for the 
scattering lengths are $1/m_{\pi}$.} \cite{Korkmaz} 
(where the first error is statistical and the second is systematic) is in good agreement with 
 the ChPT prediction of $28.2\pm0.6$ \cite{KR-chiral}. 
Using these experimental values and the relationship given above
leads to a value of 
$\beta=3.43\pm0.08$. 
On the other hand, the ChPT prediction 
(which does not satisfy unitarity) is $\beta = 2.78$ \cite{loop,loop-2,loop-3}. This difference is due to the truncation of 
the ChPT calculation at the one-loop $O(q^{4})$ level.
This value of $\beta$  is at the $\pi^+$n threshold. For
higher energies the
value is proportional to the value of the $E_{0+}(\gamma p \rightarrow \pi^+ n;E_{\gamma})$
amplitude (evaluated at photon energy E$_{\gamma}$) which is
predicted to decrease with increasing E$_{\gamma}$~\cite{DMT}.

\begin{figure}
\begin{center}
\includegraphics[width=4.0in]{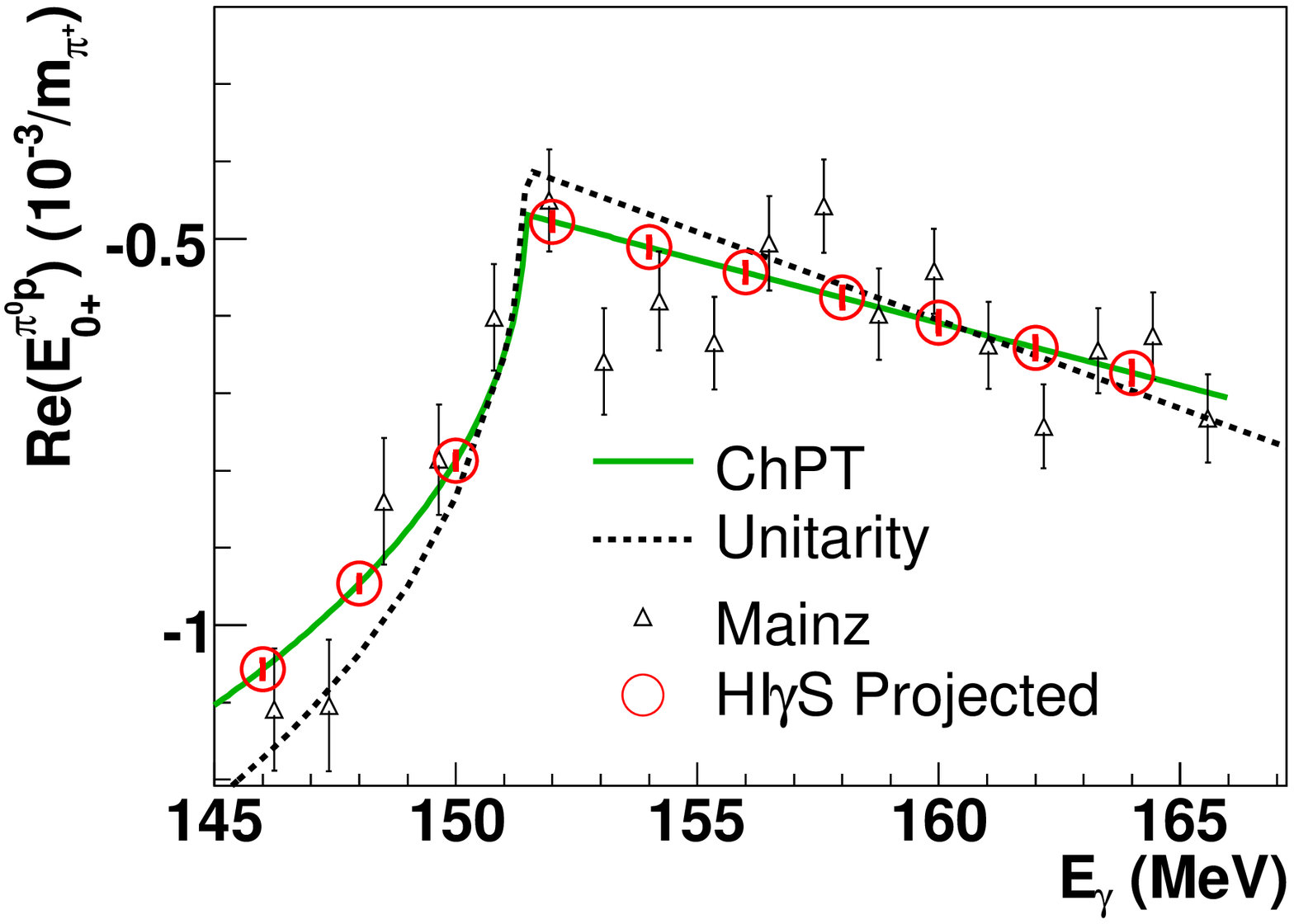}
\includegraphics[width=4.0in]{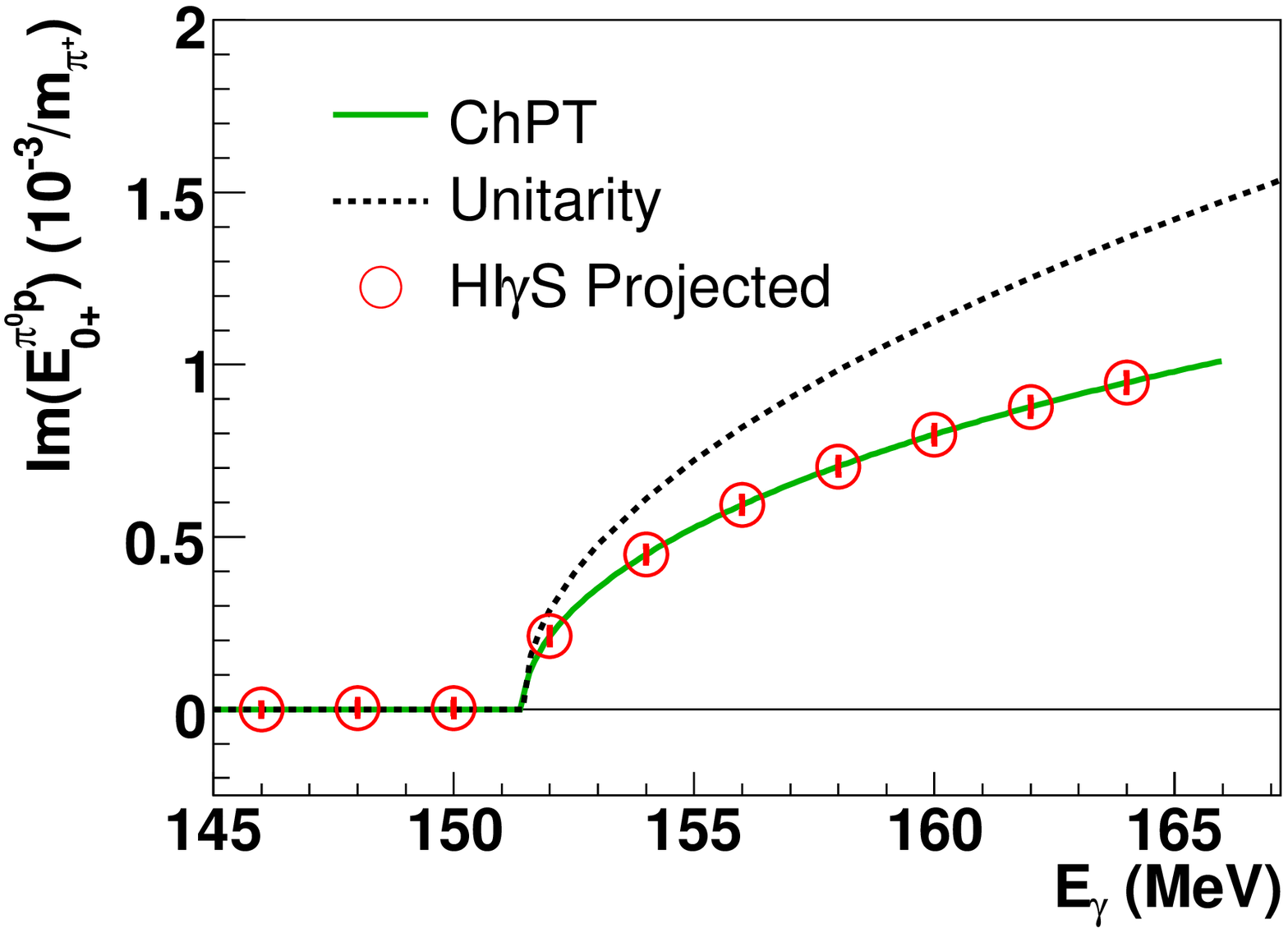}
\caption{Upper panel: Re ($E_{0+}$) for the $\gamma p \rightarrow \pi^{0}p$ reaction \cite{Schmidt}. The solid curves are  ChPT calculation \cite{loop,loop-2,loop-3, loop2}, and dashed curves are unitary fit~\cite{Mainz,Mainz-2,AB_lq} (see Eq. \ref{eq:unitary_cusp}).  Lower panel: Im ($E_{0+}$) for the $\gamma p \rightarrow \pi^{0}p$ reaction \cite{AB_lq}. For both panels the projected data points show the estimated statistical errors for 100 hours of beam time at HI$\gamma$S at each energy
in four different beam-target polarization configurations (see Fig.~\ref{fig:pi0-asym}).
\label{fig:E0p} }
\end{center}
\end{figure}

The results for Re $E_{0+}$ are presented in Fig.~\ref{fig:E0p}. The 
points are extracted from the latest data from Mainz \cite{Schmidt}. 
The data on which they are based are not a
complete experiment, which means that the multipoles cannot be extracted
without some model assumptions.
The errors shown in Fig. 1 are based on the experimental errors and do not
include any model dependence. Future experiments which contain more
polarization observables will reduce the model dependence of the analyses.
Two curves are shown in Fig.~\ref{fig:E0p}. 
They are ChPT \cite{loop2,Bernard:FFR} and a unitary fit \cite{Mainz}. On the 
basis of the data we have to date, they fit  equally well. The anticipated improvements in the determination of this multipole based on 100 hours of beam time 
per observable at HI$\gamma$S are also presented in the figure and will
be discussed in detail in Sec.~\ref{sec:future} (similar results could also be obtained at Mainz). Note that the sign of $\beta$ has been measured to be positive as predicted; a negative sign would produce a predicted curve which is above the projected linear curves. In the future, the sign of $\beta$ will be
measured since the polarized target asymmetry \textbf{$T(\theta)=A(y)$} is 
proportional to $\beta$. Thus the sign of this asymmetry is a direct measurement
of the sign of $\beta$.

At the present time there are no measurements of  Im $E_{0+}(\gamma p \rightarrow 
\pi^{0}p)$. The predictions are presented in Fig.~\ref{fig:E0p}. The discrepancy in the values of 
$\beta$ can be seen here. We anticipate that future experiments 
 will provide a measurement at the 1-to-2\% level and easily  be able to distinguish between these predictions. From the  results of the 
proposed measurement of Im $E_{0+}$ one can extract
$a_{cex}(\pi^{+}n \rightarrow \pi^{0}p)$ from the value of 
$\beta$ using the experimental results for $E_{0+}(\gamma p \rightarrow \pi^{+}n)$.   

\subsection{ Isospin Violation at Intermediate Energies}\label{sec:IS}
As was discussed in Sec.~\ref{sec:intro}, one expects isospin violations in $\pi N$ scattering to be 
$\simeq \frac{m_d -m_u}{\Lambda_{QCD}}\simeq$ 4 MeV/ 200 MeV = 2\% \cite{W2}. However, it was shown that the 
electromagnetic effects reduce this to below 1\% \cite{FM,FM-2}. Empirically,
there have been two independent claims that isospin has been 
violated in medium-energy (30 to 70 MeV) $\pi$N scattering \cite{Gibbs,Gibbs-2}. Specifically, they showed that 
the ``triangle relationship," which relates the amplitude of charge-exchange scattering ($f(\pi^-p\rightarrow\pi^0n)$) to the elastic scattering amplitudes ($f(\pi^{+(-)}p)$):
 \begin{equation}
\label{Ddef}
D \equiv f(\pi^{-}p \rightarrow \pi^{0}n)-\frac{1}{\sqrt{2} }( f(\pi^{+}p) - 
f(\pi^{-}p)) =0
\end{equation}
does not hold. Instead they found
 \begin{equation}
\label{Dexp}
\begin{array}{rl}
 D \simeq -0.012 \pm 0.003 ~\rm{fm}  \\
D/f(\pi^{-}p \rightarrow \pi^{0}n)\simeq 7 \%
\end{array} 
\end{equation}
about an order of magnitude larger than has been predicted by ChPT \cite{FM}. If true, this is a major discrepancy! However in a recent pion charge exchange experiment reported from TRIUMF~\cite{TRIUMF-cex} the results were consistent with a very small isospin breaking. The  main difference between their conclusion and those of Gibbs and Matsinos~\cite{Gibbs} is due to an experimental discrepancy between two measurements of the pion charge exchange cross section. As will be shown below it is possible to measure the effect of isospin breaking using the transverse polarized asymmetry in photo-pion production. This is an entirely different reaction and does not depend on the measurement of cross sections!

The magnitude of the isospin-breaking effects in 
electromagnetic pion production can be obtained by assuming that they occur in the s-wave.  Using Eq.~\ref{eq:M_IS} with $A_{l,j}^{2I}(\psi) = E_{0+}^{2I}(\psi)$, where I =1/2, 3/2 are two isospin states of the final $\pi$N system:
\begin{eqnarray}
\delta E_{0+}^{2I} & = & E_{0+}^{2I}(\psi)- E_{0+}^{2I}(\psi=0)\\
\mbox{which gives, for I=1/2, } \delta E_{0+}^{1} & \simeq &i \frac{\psi}{2} E_{0+}^{3} \mbox{, and }\\
\mbox{for I=3/2 } \delta E_{0+}^{3} & \simeq &i \frac{\psi}{2} E_{0+}^{1} \mbox{ .}
\end{eqnarray}
The approximation 
has been made that both the s-wave phase shifts and $\psi$ are small, which is true in this energy region. This 
approximation, as well as assuming that the
isospin breaking is in the s-wave, is easy to remove. It does serve to show  that the two isospin states are indeed mixed, that the 
isospin mixing $\rightarrow 0$ as $\psi \rightarrow 0$ and that the multipoles pick up a small imaginary part 
due to this mixing. This implies that the time-reversal-odd observables (imaginary parts of 
bilinear products of the multipoles) will have to be measured to observe this effect. This will require 
polarized target experiments. 

\begin{figure}
\begin{center}
\includegraphics[width=4.0in]{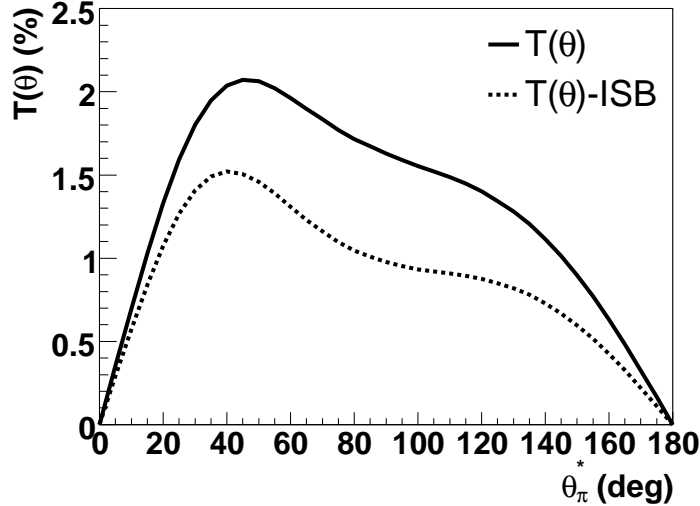}
  \caption{Polarized target asymmetry $\boldsymbol{A(y) = T(\theta)}$ in \% for the $\gamma \vec{p} \rightarrow \pi^{+}n$ reaction for 
W = 1120 MeV (E$_{\gamma}$ $\simeq$ 200 MeV). The target polarization is normal to the reaction plane. The solid curve 
is the prediction of the DMT model without isospin breaking ($T(\theta)$). The dotted curve is the DMT prediction if the 
empirical value of the isospin-breaking parameter $\psi = -0.010 \pm 0.004$ (Eq.~\ref{S_IS}) is used ($T(\theta)-ISB$)\cite{AB_lq}.\label{fig:An_plus}}
\end{center}
\end{figure}

The empirical value of the isospin-breaking parameter $\psi$
can be obtained from the measurements of D. From 
the 2x2 $\pi N$ part of the S matrix, Eq. \ref{S_IS}, one obtains \cite{AB_lq}:
\begin{equation}
D =\frac{ 1}{2q} \sin \psi  e^{(i \delta_{0,1/2}^{1} + i\delta_{0,1/2}^{3})} \mbox{ ,}
\end{equation}
where it has been assumed that the isospin violation occurs in the s-wave, and $\delta_{0,1/2}^{1}(\delta_{0,1/2}^{3})$ are the s-wave $\pi$N phase shifts in the I=1/2~(3/2) state. Note that 
$D \rightarrow 0$ when $\psi \rightarrow 0$, as expected. 
The observed consequences of isospin breaking can be calculated for  
photoproduction using the empirical value of D. The simplest such observable is the polarized-target asymmetry. An example of this is given 
in Fig.~\ref{fig:An_plus}. It can be seen that the relative effect is about twice as large in photoproduction as 
in pion scattering (15\% versus 7\%), making this a promising method to either confirm or dispute the 
reported isospin violation \cite{Gibbs,Gibbs-2}. It also should be pointed out that, in photoproduction, the final 
states ($\pi^{0}p, \pi^{+}n$) have little Coulomb interaction, which is one of the difficulties in 
analyzing pion-nucleon scattering data.
The observation that the transverse polarized itarget asymmetry can be
used to 
check isospin symmetry in the $\pi$N system is an example
of the fact that this observable is sensitive to the $\pi$N phase shifts, and
in fact can be used to measure them~\cite{AB-piN-Newsletter}.

\section{Experimental Facilities}\label{sec:facilities}
Historically, experimental studies of Chiral Dynamics in photo-pion physics
have been carried out at tagged photon facilities such as the 
Saskatchewan Accelerator Facility (SAL) \cite{SAL} and the
Mainz Microtron (MAMI) \cite{MAMI}. 
Future experiments, described in this review, are planned at the 
High Intensity $\gamma$-ray Source (HI$\gamma$S).
A brief review of the SAL and MAMI facilities and a somewhat more 
detailed description
of  the less documented HI$\gamma$S facility is presented here.
\subsection{Mainz and tagged photon facilities}
The SAL facility utilized a 300 MeV linear accelerator with a
pulse stretcher ring (PSR) to produce a $c.w.$ electron beam. 
The $\gamma$-rays were produced via bremsstrahlung in a 
narrow acceptance cone. The post-bremsstrahlung electrons were $tagged$
by a focal plane detector. The $\gamma$-ray energy was defined by
measuring the electron energy in a magnetic spectrometer and 
a time-of-flight coincidence between the tagged electron and 
the outgoing particles from the photo-induced reaction. The $\gamma$-ray
beam was unpolarized and the tagger could be operated at a maximum
rate of 1 MHz per channel~\cite{SAL}. 

At SAL, the ($\gamma,\pi^0$) measurements were 
carried out using the neutral pion spectrometer for low energy pion
photoproduction Igloo \cite{IGLOO}, constructed
from 68 lead glass Cherenkov counters. This calorimeter consisted
of four faces to form a symmetric box.  An energy resolution of 
$\sim$ 40 \% was achieved at a photon detection 
energy of 100 MeV. 
The geometry was close packed which was good for the total cross
section measurements, but suffered from poor angular resolution.
An array of liquid scintillating detectors (BC-505) 
was used to detect the outgoing neutrons in the ($\gamma,\pi^+$) reaction.
The SAL facility has since been shut down. A complete
 description of the SAL facility and experiments which were performed
at SAL can be found in \cite{NIMA324,IGLOO,Sask,PRC55,PRC58,PRC57,Korkmaz}. 

The MAMI facility uses coherent bremsstrahlung as a source of polarized 
$\gamma$-rays. Linear polarization of up to 80\% can be achieved 
at $\gamma$-ray energies above pion threshold. 
The $\gamma$-rays at MAMI are tagged like 
those at the SAL
facility. A tagging energy resolution of $\sim$ 1 to 2 MeV can be obtained 
in the
$\gamma$-ray energy range of 40 - 800 MeV. The maximum tagging rate is
$\sim$ 1 MHz per channel~\cite{NIMA368,NIMA301}.

Two different detector systems are available at MAMI for the detection of neutral
pions. The first detector system, TAPS,
consists of plastic BaF$_2$ scintillator telescopes. 
Accurate position information on the impact point
of the $\gamma$-ray can be obtained
as a result of the segmentation
of crystals in TAPS ($\Delta \sim$ 30\%
of a single crystal diameter)~\cite{NIMA346}. This array was used
for the measurement of the total cross section of the $^1$H($\gamma,\pi^0$) reaction
as well as the photon
asymmetry for $\pi^0$ production at 159.5 MeV $\gamma$-ray energy \cite{Schmidt}.
More recently, a 93 \% of 4$\pi$ NaI crystal array, Crystal Ball (CB), has been installed at MAMI.
This detector consists of 672 NaI crystals, organized in two hemispheres.
The geometric efficiency of CB is rather uniform in angle
and energy as compared to TAPS and is subject to far 
smaller systematic errors, so that both the statistical and systematic 
errors will be vastly reduced.
A new photo-pion experimental program is being planned using the crystal ball together with the TAPS detector as a forward wall, along with a central tracker. 
This combination of crystal ball and TAPS  will provide a geometrical acceptance 
close to 4$\pi$ combined with good energy and angular resolution for photopion experiments. The setup will also contain
a frozen-spin polarized target which will enable the measurements of polarization observables on the nucleons. 

\subsection{HI$\gamma$S Experimental Facility:Development and Upgrades}
The High Intensity Gamma-ray Source (HI$\gamma$S) at Duke University 
is a nearly monochromatic Compton $\gamma$-ray source 
with a very high flux, a wide energy range, and switchable polarizations.
The current HI$\gamma$S electron accelerators include a $0.18$~GeV linear accelerator 
pre-injector, a $0.18$ -- $1.2$~GeV booster injector, 
a $0.24$ -- $1.24$~GeV storage ring, and several storage ring
based Free-Electron Lasers (FELs).
At HI$\gamma$S, a high-intensity $\gamma$-ray beam is produced with a flux 
two to three orders of magnitude higher than other Compton gamma sources.
The high flux performance is realized by colliding the electron 
beam in the storage ring with a temporally matched, high-power 
FEL beam built up inside a low-loss laser resonator.
HI$\gamma$S can be operated in a wide energy range from $1$~MeV to 
about $100$~MeV in the present configuration and with future upgrades 
above $100$~MeV 
up to the pion threshold energy.
This wide energy range is the result of selecting the lasing wavelength
of the FEL and electron beam energy in the storage ring.
Highly polarized gamma-ray beams can be produced with linear and 
circular polarizations using different FEL wiggler configurations.
It is also possible to switch between the left- and right- helical 
polarizations on the minute or sub-minute time scale.
Furthermore, as a dedicated Compton gamma source, HI$\gamma$S
can be optimized for specific research programs by operating in either 
the high-flux mode or high-resolution mode with a selectable temporal 
gamma-beam structure of a continuous wave (CW) or pulsed beam.
Given these outstanding features,  HI$\gamma$S is a world-class
Compton gamma-ray source for a wide range of scientific
research programs in nuclear physics, astrophysics, medicine, and
industrial applications.

While the first HI$\gamma$S gamma beam was demonstrated 
in 1996~\cite{VLitvinenko1997}, the high performance of  HI$\gamma$S was only 
realized after a series of major upgrades of the associated accelerators and FELs 
in recent years. 
One of the major accelerator upgrades was the construction and commissioning
of a $0.18$ -- $1.2$~GeV compact synchrotron as a full-energy
booster injector operating in a periodic injection mode 
(the ``top-off'' injection).
Since its commissioning in 2006, this booster injector has enabled the 
high flux operation of HI$\gamma$S in the higher energy region 
(above $20$~MeV) in a so-called ``electron-loss mode.''
In this mode, the booster periodically refills the storage ring 
in a top-off operation to maintain the electron beam current
as Compton scattered electrons are lost at a substantial rate.
In 2005, we also brought a new (OK-5) FEL into operation
consisting of two helical wigglers which were
installed in the same straight section as the older planar OK-4 wigglers.
With this FEL upgrade, we gained the ability to produce gamma-ray beams
with both linear and circular polarizations.
The upgraded HI$\gamma$S facility is shown in Fig.~\ref{fig:HIGSFacility}.

\subsubsection{Capabilities of HI$\gamma$S in the Present Configuration}
The fully upgraded HI$\gamma$S facility began user operation in 2007 with its 
performance exceeding the design specifications for the upgrades.
The main operating parameters of HI$\gamma$S accelerators and FELs
are found in Ref.~\cite{HWeller2009}.
Since 2007, the performance of  HI$\gamma$S has been improved further in two areas.
Below $20$~MeV in the so-called ``no-loss mode'' in which the Compton scattered
electrons are retained in the storage ring, the gamma-beam flux has
been increased by a factor of two to five using a very low loss (high-finesse) FEL cavity.
The maximum total gamma-flux in this energy range exceeds $1\times10^{10}$~$\gamma$/s.
In the ``electron-loss mode,'' by increasing the maximum magnetic
field of the FEL wigglers,
the projected maximum gamma beam energy of  HI$\gamma$S is increased to about $100$~MeV 
using commercially available $190$~nm FEL mirrors. 
The updated performance of  HI$\gamma$S gamma beams in the high flux mode is 
summarized in Table~\ref{tab:HIGSPerformance}.
The gamma-beam flux available for actual experimental use depends on
collimation.
For example, for a collimated gamma-ray beam with a $3$~\% full-width
half-max energy spread, the on-target flux is about $4.5$~\% of the total
flux given in Table~\ref{tab:HIGSPerformance}.


\subsubsection{HI$\gamma$S Upgrades for $100$ -- $158$~MeV Operation}
With the current accelerator configuration of  HI$\gamma$S, the maximum 
gamma-beam energy is limited to about $100$~MeV due to the availability
of commercial FEL mirrors at the shortest wavelength of about $190$~nm.
The gamma-beam energy tuning range of HI$\gamma$S for several
FEL wavelengths are shown in Fig.~\ref{fig:GengTuning}.
Due to a limited FEL gain and the need to build up the intracavity
optical power for Compton scattering, the FEL operation requires highly 
reflective mirrors which are typically multi-layered dielectric mirrors.
The mirror damage due to intense ultraviolet (UV) and vacuum-ultraviolet
(VUV) radiation remains 
a serious problem in the wavelength regime below $190$~nm.

\begin{figure}[h]
\begin{center}
\includegraphics[width=135mm]{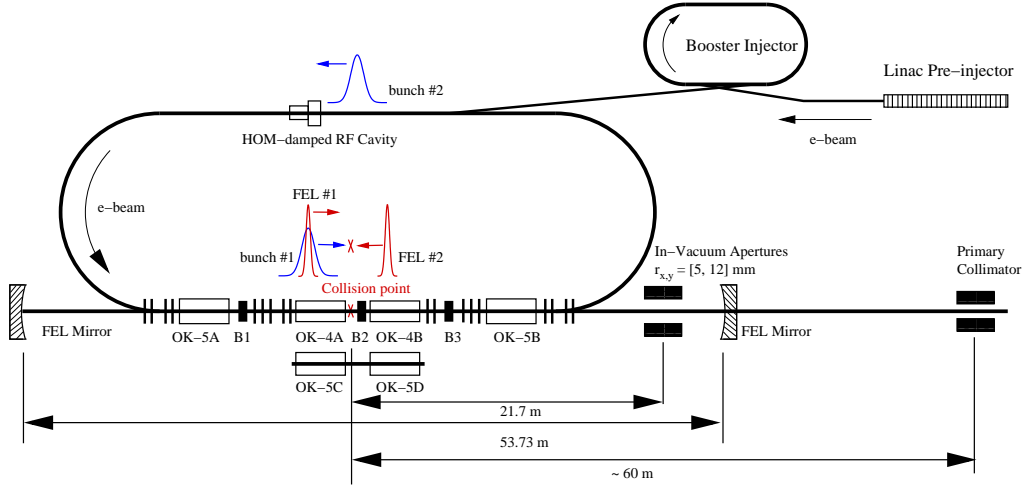}
\end{center}
\caption{
\label{fig:HIGSFacility}
The schematic of HI$\gamma$S facility in 2008 with two electron beam bunches 
colliding with the FEL pulses inside the FEL resonator cavity.
The completed upgrade projects since 2004 include
a new $0.18$ to $1.2$~GeV top-off booster synchrotron,
a new $34$~meter long storage ring straight section for hosting
a higher order mode damped RF cavity and for a new injection scheme 
with the booster, and a new OK-5 FEL in the FEL straight section.
}
\end{figure}

\begin{figure}[h]
\begin{center}
\includegraphics[width=95mm,height=140mm,angle=-90]{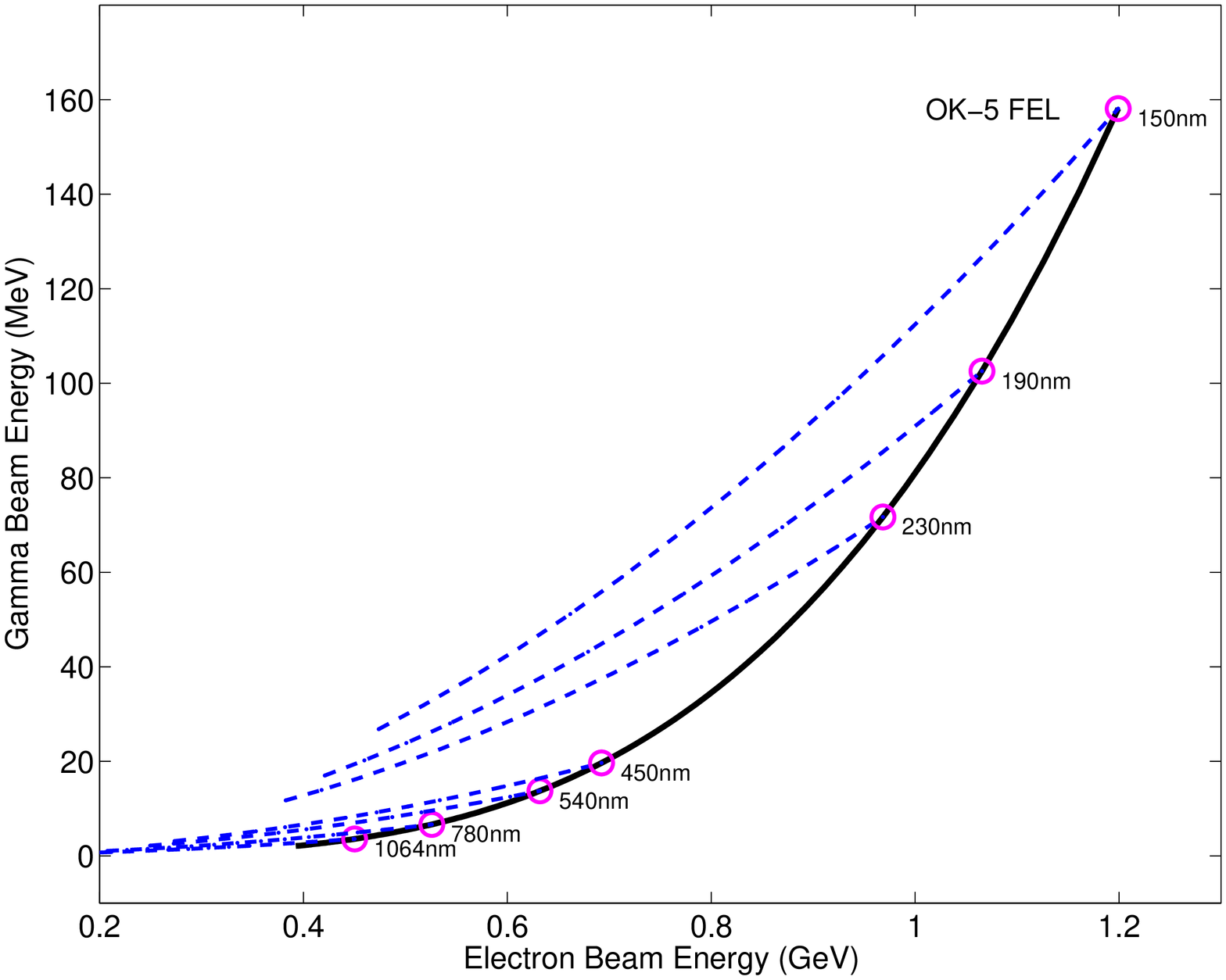}
\end{center}
\caption{
\label{fig:GengTuning}
The energy tuning range of the circularly polarized $\gamma$-ray beam with 
the OK-5 FEL from $1064$~nm down to $150$~nm.
A set of radiation resistive FEL mirrors with a high reflectivity at $150$~nm
are necessary to reach the highest gamma-beam energies
-- about $158$~MeV with the existing $1.2$~GeV storage ring.
For a given FEL wavelength, the highest gamma energy
is determined by the highest electron beam energy for FEL lasing as limited 
by the maximum magnetic field of the wigglers.
The thick black curve shows the maximum gamma energy as a function
of the electron beam energy by varying FEL wavelengths.
The blue dashed curve show the gamma energy range for a fixed
FEL wavelength.
}
\end{figure}

A critical step to extend HI$\gamma$S operation toward and beyond the 
pion-threshold energy is the development of highly reflective, 
radiation-resistive $150$~nm FEL mirrors.
This development will require advancing certain aspects
of optical coating techniques in the vacuum ultra-violent (VUV)
wavelength region.
By closely collaborating with commercial optics companies, 
we project that useful $150$~nm FEL mirrors will be available
in the next few years and we will take advantage of any new
advance with the VUV coating technology.
As another important step toward operating the FEL at $150$~nm,
two additional OK-5 wigglers will be installed by the end of 2009
to form a four-wiggler FEL (see Fig.~\ref{fig:HIGSFacility})
in order to boost the FEL gain.

With $150$~nm FEL mirrors, by operating the storage ring and booster up to
maximum energy of $1.2$~GeV,  HI$\gamma$S will be able to produce gamma-beams
at energies up to around $158$~MeV.
Like operating in other energy region in the ``electron-loss mode'',
the maximum gamma flux is expected to be limited by the electron beam 
injection rate.
The projected initial performance is $1 \times 10^8$--$2 \times 10^8$ $\gamma$/s total, and $5 \times 10^8$--$1\times 10^9$ $\gamma$/s after
increasing the electron beam injection rate by upgrading the injectors.

\subsubsection{A New Compton Gamma-source for High-flux Operation up
to $220$~MeV}

{\it Ground-breaking nuclear physics research} requires a high energy
gamma-ray source with an on-target flux of $10^9$~$\gamma$/s or higher.
This level of flux performance is now within reach with a 
specially designed Compton gamma facility which takes advantage of the 
technological advances developed at HI$\gamma$S and elsewhere.
One of the key components is the availability of a very high power
photon beam which has a temporal structure matched to that of 
the high-intensity electron beam.
At HI$\gamma$S, we have demonstrated the capability of 
producing a total gamma-flux of $10^{10}$~$\gamma$/s,
or equivalently, an on-target flux on the order of $10^{9}$~$\gamma$/s,
for few MeV gamma beams
in the ``no-loss'' mode, driven by an intense intracavity laser beam
with a kW average power in the infrared or visible region.
This level of high laser power is realized in a newly developed
very low-loss FEL laser cavity with a finesse of $> 3\times10^3$.
This development paves the way for achieving $10^9$~$\gamma$/s
gamma flux performance above $100$~MeV with a few-GeV electron accelerator.

One of the possibilities for such a high-energy,
high-flux Compton gamma source  would consist of
a $2.5$~GeV storage ring and a $480$~nm, kW FEL beam. 
Such a facility could produce gamma ray beams up to $220$~MeV.
The flux performance would mainly be limited by how fast the lost electrons
can be replaced.
For example, to produce a gamma beam with a flux of $10^9$~$\gamma$/s 
and with a $3$\% full-width, half-max energy spread, a full-energy,
top-off injector should be developed to refill the storage ring
at a rate of about $3$~nC per second. This is realizable 
with a booster injector running at a few Hz.
An even higher flux is possible if the injector and storage ring
are further optimized.
This particular configuration is rather feasible because it utilizes 
the demonstrated accelerator and FEL technologies without pushing 
new technological limits.

An alternative method for producing a high-intensity laser beam is to
use a very low loss (high-finesse) optical resonator driven by an external laser.
This technology is still under development but may become mature
enough to replace the FEL as the photon source for this future
Compton gamma source.

\begin{table}[h]
\begin{center}
\caption{
\label{tab:HIGSPerformance}
\small
Parameters of HI$\gamma$S gamma-ray beam in the high-flux mode (2008).
}
\vspace{0.1in}
\begin{scriptsize}
\begin{tabular}{||l|c|c||}      \hline
Parameter                         & Value             & Comments    \\ \hline
E-beam configuration              & Symmetric two-bunch &\\
E-beam current [mA]               & $50$ - $120$      & Total current\\\hline
$\gamma$-ray energy, $E_\gamma$ [MeV]       &                   & \\
\fivesp With mirrors $1064$ to $190$~nm  
                                  & $1$ -- $100$      & With existing hardware \\ \hline
Total flux [$\gamma$/s]                & & \\
(a) No-loss mode ($\leq 20$~MeV)     & & Both linear and circular \\
\fivesp $E_\gamma = 1$ -- $2$~MeV      & $1\times10^8$ -- $1\times10^9$$^{(a)}$ & polarization\\   
\fivesp $E_\gamma = 2$ -- $20$~MeV     & $1\times10^9$ -- $1\times10^{10}$    & \\   
(b) Loss mode ($> 20$~MeV)        & & Preferred Polarization\\
\fivesp $E_\gamma = 21$ -- $100$~MeV   &  $1\times 10^8$ -- $2\times10^8$$^{(b)}$  & Circular \\ \hline
Linear and Circular polarization  & $>95$\% & Depending on collimator size\\\hline
\end{tabular}
\end{scriptsize}
\end{center}
\begin{scriptsize}
$^{(a)}$ High flux horizontally polarized gamma-ray beams can be produced 
by the OK-4 FEL. The circularly polarized gamma-ray flux is lower due to
the dynamic impact of the OK-5 wigglers. \\
$^{(b)}$ The flux is currently limited by the capability of sustaining
a high intracavity power by the FEL mirrors and the electron injection rate.\\
\end{scriptsize}
\end{table}
\vspace{0.1in}

\subsection{Experimental Setup at HI$\gamma$S}

The experimental setup for the proposed experiments at HI$\gamma$S includes the following
key resources : 1) Crystal Box detector~\cite{XTALBOX}; 2) Neutral Meson Spectrometer (NMS)~\cite{NMS}; 3) 
Liquid Scintillating Neutron Detectors; and 4) HI$\gamma$S Frozen-Spin Polarized Target (HIFROST).
The Crystal Box detector consists of 270 NaI crystals arranged in three arrays of 9 $\times$ 10
crystals each. The size of each crystal is 2.5 in. $\times$ 2.5 in across and 12 radiation lengths long.
The energy resolution $\Delta$E/E (FWHM) 
of each array is $\sim$ 1.3 \% at 100 MeV of photon detection energy. The NMS consists of 120 CsI crystals
arranged in 2 arrays of 6 $\times$ 10 crystals each and $\sim$ 16 radiation
lengths long.  A detailed description of the Crystal Box and the NMS detectors
can be found in \cite{XTALBOX,NMS}. In addition, a variety of Liquid Scintillating Neutron Detectors are
available at TUNL/HI$\gamma$S for the detection of neutrons from the proposed  
$^1$H($\gamma,\pi^+$)n reaction. The most prominent of these detectors is the Blowfish array, which is
a reconfigured version of the detectors used at SAL for the cross section measurement of the $^1$H($\gamma,\pi^+$)n 
reaction. Therefore, it is ideally suited for a similar measurement at HI$\gamma$S with only minor modifications.
A detailed description of the Blowfish array is available in \cite{Blackston} and references therein.

A frozen-spin polarized target (HIFROST) is being installed at HI$\gamma$S.
The HI$\gamma$S target has a length of 7.3 cm, a
diameter of 2.0 cm, and a target thickness of
$\sim$ 3.5 $\times$ 10$^{23}$ polarized protons per cm$^2$.
The polarization is expected to be 70-80 \%. The target is first polarized using
 a strong (4 T) field in a superconducting magnet and the polarization is maintained
using superconducting  $holding$ coils at a lower ( $\sim$ 0.24 T) field. These holding coils can be
cylindrical to produce longitudinal polarization along the axis of the cylindrical
target, or a pair of isolated $saddle$-shaped coils to provide polarization transverse to the axis
of the cylindrical target. The longitudinal coils are a proven technology and have been used
at Paul Sherrer Institute (PSI), MAMI, and JLab \cite{FST,MAMIT}. The $saddle$ coils are being developed in
collaboration between TUNL and JLab. Recently, multiple $saddle$ coils were prepared at JLab and
tested for their maximum field strength. A Similar setup at TUNL is now ready to produce and test
these saddle coils. Recent results indicate that the current design and production scheme of the saddle coils
produces the desired field for the transverse polarized target. In addition, the collaboration
is developing a $scintillating$ frozen spin polarized target. 
The U.Va.--TUNL group in collaboration with the University of Massachusetts group 
(led by Rory Miskimen) and James Madison University (led by S. Whisnant), is planning to build such a scintillating-frozen-spin target.
This target will be based on the experience of the
group at the PSI which has constructed a polarized scintillating frozen-spin
target \cite{PSI}. We expect the non-scintillating HIFROST to be fully installed at HI$\gamma$S by the summer of 2009, and the
scintillating version available by mid-2010.

The $\pi^0$ from the $^1$H($\gamma,\pi^0$) reaction will be detected primarily by the Crystal Box with possible additional
coverage by the NMS. 
One possible setup is placement of the three crystal box arrays
in a triangular configuration, with their long edges touching one another and parallel to the beam direction. In this setup, the
target is placed in the center of the triangle with the shortest distance from the target to each crystal box array being 6.5 in. At E$_{\gamma}$ of 
158 MeV, this setup provides $\sim$ 3$\pi$ solid angle coverage for $\pi^0$ detection. This could be enhanced to over $\sim$ 90\%
of 4$\pi$ with the inclusion of the NMS in the setup. The energy of the neutral pion is inferred from its decay photons. 

We expect a flux of $\sim$ 1 $\times$ 10$^8$ $\gamma$/s on target with an energy spread of $\sim$ 5 \%. 
Since the $\gamma$-rays are produced using two electron bunches, they are produced 
in a pulsed mode with a repetition rate of 5.58 MHz (a beam burst every $\sim$ 179 ns). This pulsed structure
of the beam helps in making the $time-of-flight$ technique a valuable resource in reducing non-beam related background. 
The $\gamma$-ray flux will be measured using a large NaI detector of known efficiency. This detector is placed directly in
the beam with the flux reduced by inserting precision attenuators in the beam. Knowing the attenuation coefficients and the
number of attenuators, a measurement of the total unattenuated flux can be
obtained and used to calibrate a thin scintillating
paddle. Monitoring the flux with this thin-scintillator provides a real-time flux determination. This
attenuator system exists at HI$\gamma$S and has been employed and tested to provide flux measurements to 
an accuracy of 3 to 7 \%.

Details of the expected count rates and extraction of multipole moments to
construct the desired observables is given in Sect. \ref{sec:proton}. 

\section{Photoproduction Observables with Full Polarization} \label{sec:formulas}
\subsection{Observables}\label{sec:obs}

The differential cross section for pion photoproduction in the center of mass system has the form~\cite{MAID-eta}: 
\begin{eqnarray}
d \sigma/d \Omega_{\pi} & = & (p_\pi^*/k_\gamma^*)
\left\{ R^{00}_T + \Pi_T { }^c \! R^{00}_{TT} \cos 2 \varphi \right.
+ P_x \left( - \Pi_T { }\boldsymbol{^s \! R_{TT}^{0x}} \sin 2 \varphi + \Pi_{\odot} R_{TT'}^{0x} \right)
\nonumber \\
& & +P_y \left( \boldsymbol{R_{T}^{0y}} + \Pi_T\boldsymbol{ { }^c \! R_{TT}^{0y}} \cos 2 \varphi \right)
\left. + P_z \left(- \Pi_T\boldsymbol{ { }^s \! R_{TT}^{0z}}\sin 2 \varphi + \Pi_{\odot} R_{TT'}^{0z} \right)
\right\}
\;
\label{eq:cross}
\end{eqnarray}
 where each R is a response function (bold-face response functions indicate
a time reversal odd function), $p_\pi^*$ and $k_\gamma^*$ are the center of mass momenta of the pion and photon, respectively, $\varphi$ is the angle between the photon polarization vector and the reaction plane, $\Pi_T$ and $\Pi_{\odot}$ are the magnitudes of the photon's linear and circular polarization, respectively, on a scale from zero to one. The $P_{x,y,z}$ terms are the  magnitudes of the target polarization in the $\pi N$ center of mass reaction plane where $z$ is the  photon direction, $x$ is transverse in the reaction plane, and $y$ is transverse and perpendicular to the reaction plane. $R_{T}^{00}$ is the unpolarized observable and ${ }^c \! R^{00}_{TT}$ and $R_{T}^{0y}$ are the single polarization observables. The notation denotes each response function for three different polarized target directions: $R^{0x}, R^{0y}$,and $R^{0z}$. The four boldface symbols represent time reversal odd observables (TRO; imaginary parts of the bilinear products of the multipoles). The other four response functions are time reversal even (TRE; real parts of the bilinear products of the multipoles). The TRO amplitudes are crucial in measuring the phases of the amplitudes, which makes them more sensitive to the $\pi N$ physics.  The different terms in Eq.~\ref{eq:cross} can be separated by their dependence on $\varphi$, the target polarization directions, and by varying the sign of the linear and circular beam polarizations. 
 
 Equation~\ref{eq:cross} is in the CM system so we must Lorentz transform the observables from  the laboratory system: this is taken  as a right handed coordinate system in which the photon direction is $z_{L}$ = $z$, $x_{L}$ is transverse in the plane of the floor, and $y_{L}$ is transverse and perpendicular to the plane of the floor. In this lab system we define the pion direction as $\theta_{\pi}^{L}, \phi_{\pi}$, so  that the orientation of the reaction plane is $\phi_{\pi}$ (the value of $\phi$ is the same in the lab and the CM system).  When the photon polarization is chosen parallel to $x_{L}(y_{L}), \varphi=\phi_{\pi}(\pi/2 -\phi_{\pi})$. The value of $\theta_{\pi}$ in the CM system is obtained from $\tan(\theta_{\pi}) = \beta_{\pi}\sin(\theta_{\pi})/[\gamma_{CM}(\beta_{\pi}\cos(\theta_{\pi})-\beta_{CM})]$ where $\beta_{\pi} = p_{\pi}/E_{\pi}, \beta_{CM}= k/(k+M_{p}), \gamma_{CM} = (1-\beta_{CM}^{2})^{-1/2}$  and k is the photon energy in the lab system. We know the target polarization in  the lab system where  it has  
magnitude $p_{T}$, points in the  direction $\theta_{T}, \phi_{T}$ and  its four vector is (1, $\vec{P_{T}})$. 
 To obtain the target polarization in the CM system we have to perform another Lorentz transformation. As a consequence,  the $z$ component is transformed.  
The transverse target polarization components remain the same in magnitude as in the lab system
 $P_{xL}  = P_{T}\sin(\theta_{T})\cos(\phi_{T}), P_{yL}=P_{T} \sin(\theta_{T}) \sin(\phi_{T})$. However care must be taken in calculating $P_{x}, P_{y}$ since these coordinates are defined relative to the pion production plane which is at angle $\phi_{\pi}$ relative to the lab plane. In this frame the $y$ axis (normal to the reaction plane) is in the direction of $\hat{p_{\pi}^{*}} \times \hat{k}$ and
  $\hat{x} =\hat{y}\times \hat{k}$ where $p_{\pi}^{*}$ is the pion center-of-mass (CM) momentum. Therefore the target polarization in the CM frame is
  \begin{eqnarray}
P_{x}  = P_{T} \sin \theta_{T} \cos(\phi_{T} - \phi_{\pi}), \\ \nonumber
P_{y} = P_{T} \sin \theta_{T} \sin(\phi_{T} - \phi_{\pi}), \\ \nonumber
P_{z} = \gamma_{CM}(p_{T}\cos(\theta_{T})-\beta_{CM})
\label{eq:PolT}
\end{eqnarray} 
 Note that if $\theta_{T} = 0$ the CM polarization is longitudinal, but if $\theta_{T} = 90^{\circ}$ there is still a longitudinal target polarization component in the CM system. In order to have a transversely polarized target in the CM system one has to choose 
 $cos(\theta_{T}) = \beta_{CM}/P_{T}$. As an example for $p_{T}$ =0.90 and a photon energy of 155 MeV, $\theta_{T}= 80.9^{0}$, a significant deviation from  $\theta_{T}= 90^{\circ}$. As can be seen from Eq.~\ref{eq:PolT} it is possible to produce a longitudinal target by pointing the target polarization in the beam direction. For the transverse polarization components another $\phi_{\pi}$ dependence is introduced. 

Experimentally one usually measures asymmetries, which are generally obtained with more accuracy since they are less sensitive to normalizations than absolute cross sections. The seven asymmetries that can be obtained from the eight response functions are presented in Table~\ref{table_obs}. Here we introduce a new, more transparent notation as well as giving the historical one. 

\begin{table}
\begin{center}
\begin{tabular}{|ccc|l|}
\hline
Observable  & & Response Function & Name \\
\hline
$\sigma_{T}(\theta^*_{\pi})$ & = & $(p_{\pi}^{*}/p_{\gamma}^{*})R^{00}_T$            & Transverse Differential Cross Section\\
\hline
$A(\vec{\gamma})\equiv \Sigma(\theta)$ & $=$ & $ -R^{00}_{TT}/R^{00}_T$   & Polarized Photon Asymmetry\\
$\boldsymbol{A(y)\equiv T(\theta)}$      & = & $  \boldsymbol{R_T^{0y} /R^{00}_T}$   & Polarized Target Asymmetry\\
\hline
 $A(\gamma_{c},z)\equiv E(\theta)$ & = & $-R_{TT'}^{0z}/R^{00}_T$ & Circ. Photon-  Long. Target\\
$A(\gamma_{c},x) \equiv F(\theta)$ & = & $ R_{TT'}^{0x}/R^{00}_T$ & Circ. Photon - Trans. Target\\
 $\boldsymbol{A(\vec{\gamma},z)\equiv G(\theta)}$ & = & $\boldsymbol{-R_{TT}^{0z}/R^{00}_T}$  & Trans. Photon - Long. Target\\
 $\boldsymbol{A(\vec{\gamma},x)\equiv H(\theta)}$ & = & $ \boldsymbol{R_{TT}^{0x}/R^{00}_T}$  & Trans. Photon - Trans. Target\\
$\boldsymbol{A(\vec{\gamma},y) \equiv P(\theta)}$      & =& $ \boldsymbol{-R_{TT}^{0y}/R^{00}_T}$ & Trans. Photon - Normal Target\\
\hline
\end{tabular}
\end{center}
\caption{Observables for photo-pion production~\cite{MAID-eta} with polarized photons and
targets. The top line is the unpolarized cross section. All other entries
are asymmetries. The second box contains the single polarization observables. The
next group consists of the double polarization observables. The last entry can also
be observed as the recoil polarization asymmetry. A new, more transparent
notation is introduced here for the asymmetries
(e.g. $\boldsymbol{A(\vec{\gamma},y)}$), and the historical notation
(e.g. $\boldsymbol{ P(\theta)}$) is also presented. As throughout this proposal, the
time-reversal-odd observables (imaginary bilinear combinations of multipoles) are
indicated with boldface.  Only six of the eight observables are
independent~\cite{obs,obs-2}.
\label{table_obs}
 }
\end{table}

There are eight response functions in Eq. \ref{eq:cross}  but only six of them are independent\cite{obs,obs-2}. At each pion emission angle there are four complex invariant amplitudes\cite{obs,obs-2}. Taking into account the fact that one overall phase is irrelevant, this makes  seven real numbers  to determine experimentally in a complete experiment (i.e one in which all of the amplitudes are determined experimentally). Since an experiment with fully polarized photons and targets can only determine six of them, one must perform at least one more experiment in which the polarization of the recoil nucleon is measured in order to perform a complete experiment\cite{obs,obs-2}. 

There is also the important issue of how many partial waves are contributing. In the low energy regime one usually assumes that only the  s and p wave multipoles are important. If one assumes this, then the angular distributions of all of the observables are  limited; e.g. $\sigma_{T} = A_{T}  + B_{T} \cos\theta +C_{T} \cos^{2}\theta$. If d waves become sufficiently important then $\sigma_{T} = A_{T}  + B_{T} \cos\theta +C_{T} \cos^{2}\theta +D_{T} \cos^{3}\theta$. In addition to the greater angular variation when d waves are important, the numerical values of  $A_{T},B_{T}, C_{T}$ are not the same. This simple example indicates the need to cover a sufficient angular range in each experiment and also sets requirements for the angular resolution and binning of the experiments. In general one must make reasonable (but model dependent) estimates of the contributions of the higher partial waves which contribute in order to make sure that significant contributions are not being overlooked. 

We shall conclude this section by giving an example of how the multipoles have been extracted from the data at low energies. In the low energy region (below $\simeq$ 165 MeV) it has been assumed that only s and p wave multipoles contribute and furthermore that the three p wave multipoles are purely real (see e.g.\cite{Mainz,Mainz-2,Schmidt}). In this approximation (which must be checked with more accurate, future data)  five numbers must be determined, $Re E_{0+}, Im E_{0+}, P_{1}, P_{2}$, and $P_{3}$, where these three P-wave amplitudes are defined in terms
of the usual multipole amplitudes by:
\begin{eqnarray}
P_1 & = & 3E_{1+} + M_{1+} - M_{1-} \nonumber \\
P_2 & = & 3E_{1+} - M_{1+} + M_{1-} \nonumber \\
P_3 & = & 2M_{1+} + M_{1-} \mbox{ ,}
\end{eqnarray}
where M$_{1+}$ and M$_{1-}$ are the P-wave magnetic dipole amplitudes
for $j = \frac{3}{2}$ and $\frac{1}{2}$, respectively, and 
 E$_{1+}$ is the P-wave electric quadrapole amplitude with $j = \frac{3}{2}$
\cite{PRC57}.
 From the measurement of $\sigma_{T}$ one obtains
$A_{T} = \mid E_{0+}\mid^{2} 
+ 1/2(\mid P_{2} \mid^{2}+ \mid P_{3} \mid^{2}), 
B_{T}= 2 Re(E_{0+} P_{1}^{*})$, and $ C_{T} =\mid P_{1} \mid^{2} - 1/2(\mid P_{2} \mid^{2}+ \mid P_{3} \mid^{2})$. The next step  was taken at Mainz by the measurement of the polarized photon asymmetry\cite{Schmidt} which determines 
$ R_{TT}^{00}= 1/2(\mid P_{2} \mid^{2}- \mid P_{3} \mid^{2})\sin(\theta)^{2}$. This still leaves us one observable short of an experimental determination of all of the multipoles, even with the restricted assumptions. The observable of choice is the time reversal odd polarized target asymmetry $\bf{R_{T}^{0y}}$ = $Im[E_{0+}(P_{3}-P_{2})]\sin(\theta)$ (assuming real P-amplitudes) from which we can obtain $Im E_{0+}$. It should be noted that if there is a contribution from higher partial waves or if the imaginary parts of the p wave multipoles are not negligible then there will be  additional $\theta$ dependent terms in $\bf{R_{T}^{0y}}$. This can be experimentally tested with new data. 

There are also two time reversal even asymmetries which  are very large in the threshold region which can provide additional and precise measurements of the multipoles which will further test the assumptions on which the present data are analyzed. There are the two  double polarization asymmetries induced with circular polarized photons 
 $A(\gamma_{c},z)\equiv E(\theta)$ and $A(\gamma_{c},x) \equiv F(\theta)$. The numerator of the latter is $ R_{TT^{'}}^{0x} = \sin \theta Re [ (E_{0+}^{*} + \cos\theta  P_{1}^{*})(P_{2}-P_{3})]$. It can be seen that a measurement of the angle ($\theta_{0}$, not 0 or 180$^{\circ}$) for which this observable = 0 provides an independent determination of $Re (E_{0+}^{*}) = - \cos\theta_{0} Re( P_{1})$. The full formulas for the observables are provided in the appendix. 

\section{Photo-Pion Production From the Nucleon}\label{sec:proton}
\subsection{Previous $\gamma p \rightarrow \pi^{0} p$ Experiments: Comparison With Theory}\label{sec:previous}
The  observables that are most sensitive to the spontaneous hiding (breaking) of chiral symmetry in QCD are those that vanish in the chiral limit (light quark masses, or $m_{\pi} \rightarrow$ 0). In $\pi$N physics these include the $a(\pi N)$: the s-wave $\pi$N scattering and charge exchange scattering lengths~\cite{W1,W2}, and $E_{0+} (\gamma^{*} N \rightarrow \pi^{0} N$): the electric dipole amplitude for electromagnetic $\pi^{0}$ production on the proton and neutron for real and virtual photons~\cite{loop,loop-2,loop-3,loop2}. In these cases the entire amplitude arises from the contributions of the small but finite quark (pion) masses and their momenta (assumed to be small) calculated in ChPT~\cite{ChPT-review,ChPT-Baryons}. A comparison of theory and experiment constitutes a low energy test of QCD and its symmetry properties, assuming the accuracy of ChPT. In the near future we expect that  lattice calculations will begin to play a role. 
Therefore recent experiments have focused on  observables that  
vanish in the chiral limit~\cite{AB_workshop}. In the case of photo-pion production this means the $\gamma p \rightarrow \pi^{0}p$ reaction. In the future we anticipate that with the use of deuteron targets this will also include the accurate measurement of the $\gamma n \rightarrow \pi^{0} n$ reaction. As will be shown, the experiments for the unpolarized cross sections have reached a reasonable state of accuracy and there is good agreement between the one loop ChPT calculations and experiment. However further accuracy in the measurements of the unpolarized cross sections are anticipated in the near future and we are just at the beginning of experiments making extensive use of polarization degrees of freedom. These will provide far more stringent tests of the ChPT calculations. 

\begin{figure}
\begin{center}
\epsfig{file=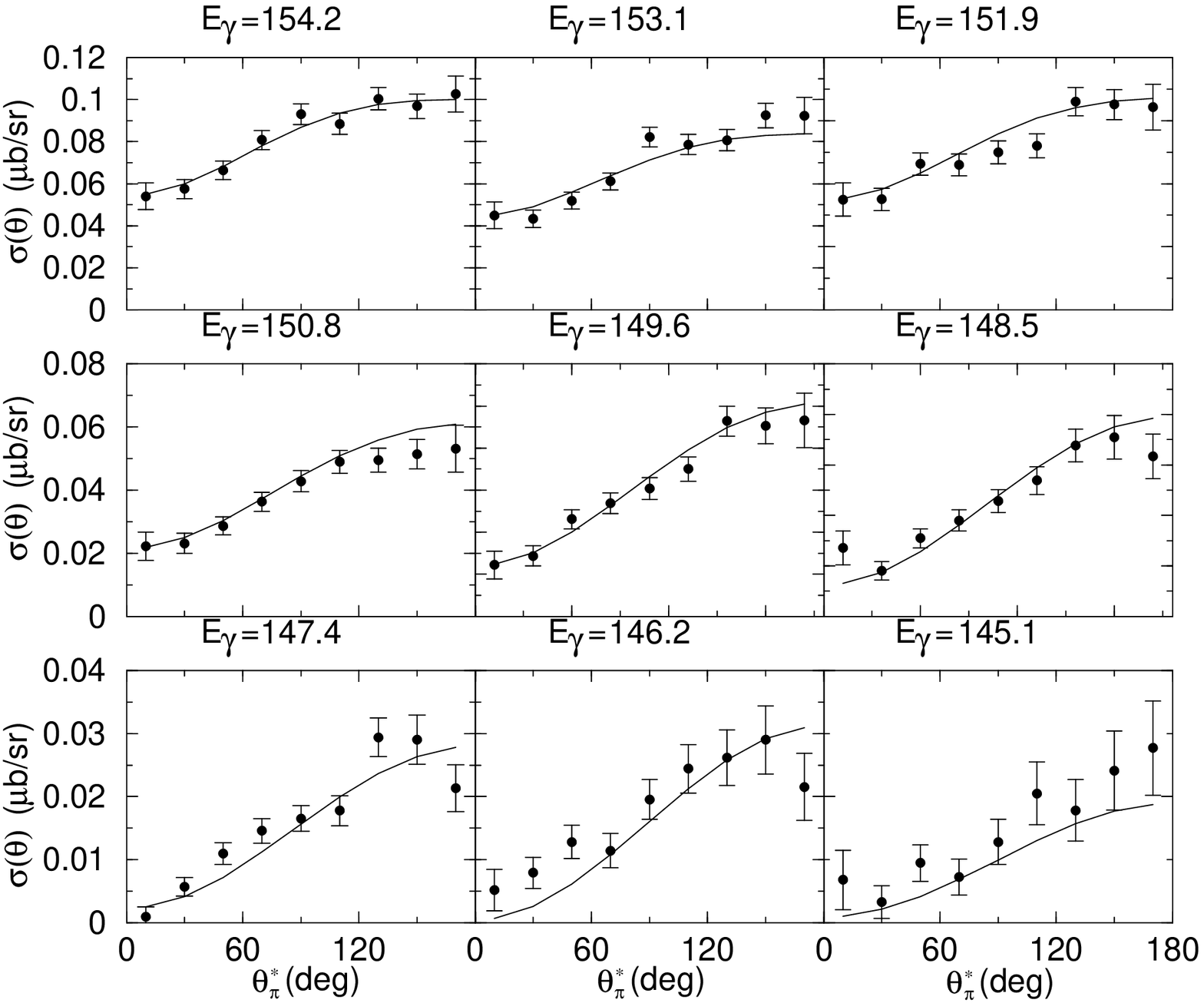,width = 12cm}
 \caption{Experimental differential cross sections in $\mu$b/sr versus pion center-of-mass angle for the 
$\gamma p \rightarrow \pi^{0}p$ reaction for a series of photon energies, labeled above each figure, starting 
just above threshold (144.7 MeV)~\cite{Schmidt}. The errors are statistical and do not include the 5~\% systematic error. The curves are the latest O($p^{4}$) ChPT calculation~\cite{loop2} 
fit to the data (see text for discussion).}
\label{fig:dsig_Mainz}
\end{center}
\end{figure}

Threshold photoproduction experiments with high-duty-cycle 
accelerators have been carried out at Mainz~\cite{Mainz, Schmidt} and 
Saskatoon~\cite{Sask}. In general, there is very reasonable agreement between the Mainz and Saskatoon 
data, and, for brevity, only the latest Mainz results~\cite{Schmidt} will be presented. These data cover 
the energy region from just above threshold (144.7 MeV) to 166 MeV in $\simeq$ 1 ~MeV bins with tagged  photons. The cross-section results for the energy region from threshold to 154 MeV 
are shown in Fig.~\ref{fig:dsig_Mainz}. The agreement with the O($p^{4}$) ChPT curves~\cite{loop2} is excellent 
and will be discussed below. The agreement of the ChPT calculations with the differential cross section data up to 166 MeV is equally good. 

\begin{figure}
\begin{center}
\epsfig{file=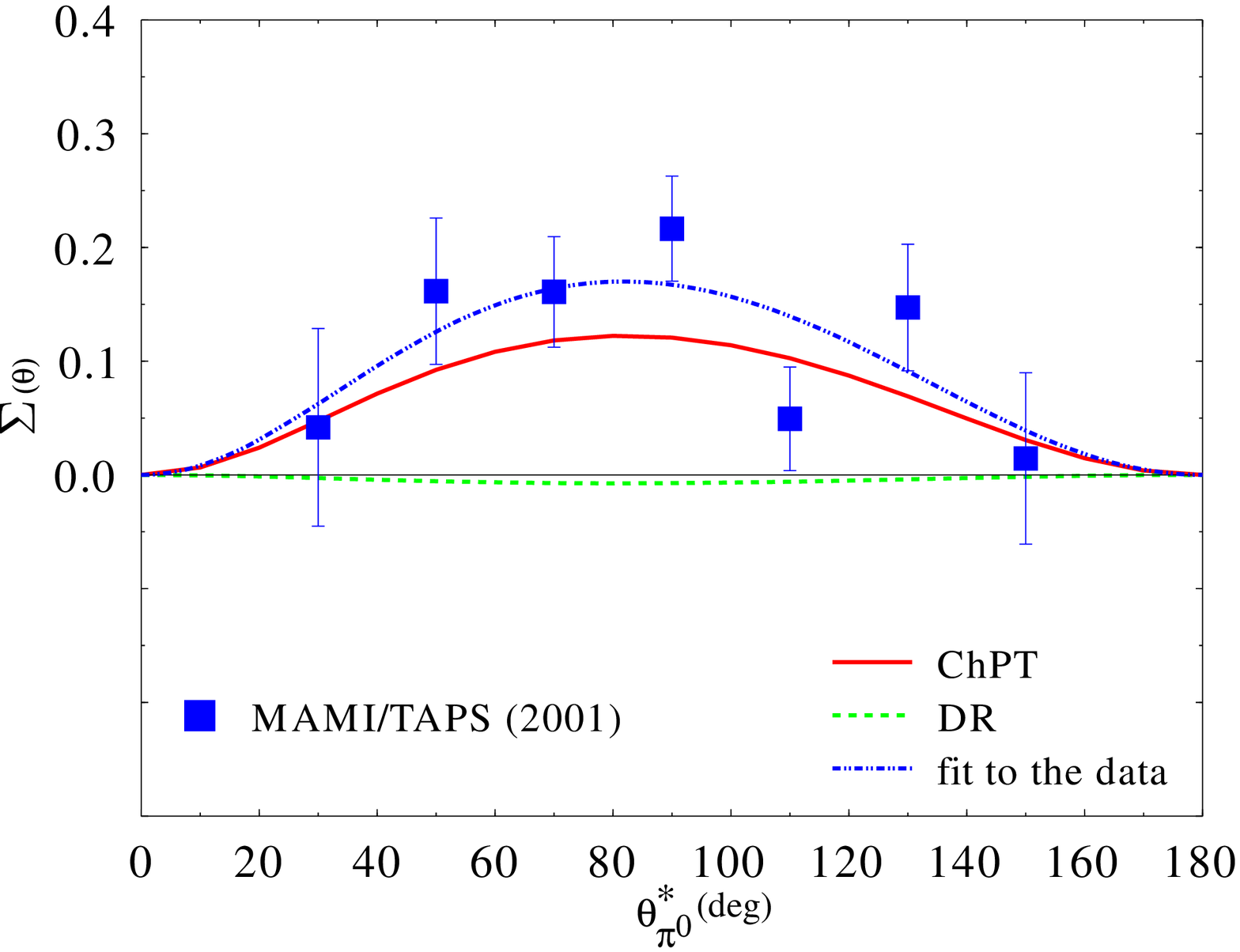,width = 12cm}
\caption{Experimental polarized photon asymmetry ($A(\vec{\gamma}) \equiv \Sigma(\theta)$) versus pion center-of-mass 
angle for the $\gamma p \rightarrow \pi^{0}p$ reaction.These data are the cross section weighted asymmetries from threshold to 166 MeV  at the average energy of 159.5 MeV~\cite{Schmidt}. The errors are statistical and do not include the 3\% systematic error. The 
solid curve, labeled ``ChPT,'' is an older [O($p^{3})$] ChPT calculation~\cite{loop,loop-2,loop-3}. The  curve with the small dots, 
labeled ``fit to the data,'' is a fit to the data and also is the same as the newer [O($p^{4})$] ChPT calculation~\cite{loop2} with several new  
low energy parameters.}
\label{fig:Sigma_Mainz}
\end{center}
\end{figure}

The latest Mainz data~\cite{Schmidt} include the first use of linearly polarized photons in the threshold region for 
the $\gamma p \rightarrow \pi^{0}p$ reaction. For statistical reasons, the entire energy range from threshold to 
165 MeV was combined at the cross-section weighted average energy of 159.5 MeV. These data are shown in Fig.~\ref{fig:Sigma_Mainz}.   
There are two ChPT calculations that are being compared to the experiments. The older ChPT calculation~\cite{loop,loop-2,loop-3} 
was carried out to O($p^{4}$) in the s-wave and to O($p^{3}$) in the p-wave multipoles. It has three 
empirically-determined counter terms, two in the s-wave and one in the p-wave multipoles. These were fit to 
the older Mainz and Saskatoon data~\cite{Mainz,Mainz-2,Sask}, which only measured the unpolarized cross section. This 
meant that Re$ E_{0+}$ and two linear combinations of the p-wave multipoles were determined. The newer 
Mainz data~\cite{Schmidt} contain one average measurement of the linear photon asymmetry, which 
determines the third p-wave multipole at 159.5 MeV. Following this measurement, the ChPT calculations  
improved by also bringing the p-wave calculation to O($p^{4}$), 
consistent with the older s-wave calculation~\cite{loop2}; 
this marks the completion of the full one-loop calculation. In this calculation, there are now two more p-wave 
low-energy parameters which must be determined empirically. This was done by fitting to the new Mainz data. The resulting calculation gives the same curve as the empirical fit to the data( labeled fit). The improvement between the older and newer 
ChPT is also easily seen. It should be pointed out that the measurement of 
$A(\vec{\gamma}) \equiv \Sigma(\theta)$ is a very sensitive test of the theoretical calculations. Most models give the wrong sign for this quantity. Even dispersion relations predictions are not in agreement with the data (Fig.~\ref{fig:Sigma_Mainz}) indicating a deficiency in the data base on which this prediction is based. 

At this point, there is impressive agreement between the data and the new,
full, one-loop O($p^{4}$) ChPT
calculations \cite{loop2}( see
Figs.~\ref{fig:E0p},~\ref{fig:dsig_Mainz}, and ~\ref{fig:Sigma_Mainz}).
However
concerns about  how well the ChPT series is converging have been expressed
for $E_{0+}$\cite{loop}. At O($p^{4}$) the p wave multipoles also showed
that there were some significant effects ($\simeq$ 25\% ) which are almost
cancelled out by a $\Delta$ contribution.
These calculations were performed in the framework of  Heavy Baryon ChPT
(HBChPT) \cite{loop,loop2}. Some open theoretical issues include the use of
the relativistic ChPT\cite{Becher}, which  might provide superior
convergence. In addition there is the issue of the inclusion of the $\Delta$
as an active degree of freedom in ChPT calculations. To effectively carry
out these calculations to higher order it may be useful to combine ChPT with
dispersion relations. Some starts in this direction have been made. For
example the most accurate calculation for Re $E_{0+}$ have been made with
relativistic ChPT combined with the Fubini-Furlan-Rosetti sum
rule~\cite{Bernard:FFR}.
Finally, there is the  issue of the upper limit to which this effective
field theory is accurate. It is important to have  experiments to explore
the limits and see at what point theory and
experiment diverge. For all of these reasons, it is important to extend the
accurate tests of the theory to include more sensitive polarization
observables  which are  more sensitive to the details of the theoretical
predictions. Plans for the future will be discussed in Sec.\ref{sec:future}.

In addition to ChPT calculations there are  many  models of pion
photoproduction, including MAID\cite{MAID}, a unitary-isobar model
that is primarily designed to provide flexible fits of the data in the
$\Delta$ and $N^{*}$ regions, as well as other dynamical models~\cite{Sato,
FernandezRamirez}.
However, in the near-threshold region,  MAID or the other dynamical model
calculations do not agree very well with the existing data. The DMT
(Dubna-Mainz-Tapei)
dynamical model is an extension of MAID and has proven to be more accurate
in the near-threshold region~\cite{DMT}.
This can be used to estimate the magnitude of the $\Delta$ contribution
which increases rapidly with photon energy and provides a planning tool
for  projected experiments and for comparison with future data.
In addition there is the empirical SAID multipole analysis of the existing
data~\cite{SAID}.

\subsection{Future $\vec{\gamma} \vec{p} \rightarrow \pi^{0}p $ Experiments}\label{sec:future}

In this section we present the physics that can be obtained from precise measurements of  photo-pion production from the proton from threshold to intermediate energies below the $\Delta$ resonance. 
The emphasis is on the open questions, keeping in mind the experimental capabilities that presently exist or are in an active stage of development. 
In Section~\ref{sec:facilities}, we have presented an overview of the unique capabilities at HI$\gamma$S and Mainz
to perform  $\vec{\gamma} \vec{p} \rightarrow \pi^{0}p,~\pi^{+}n$ experiments with beam and target polarizations. These two facilities provide complementary capabilities. At Mainz the experiments will be carried out with 
tagged-photons in which  a range of photon energies and pion angles will be measured simultaneously. At HI$\gamma$S, the plan is to 
measure both charge channels at several photon energies with high statistics.

The data up to a photon energy of $\simeq$ 165 MeV will provide a 
stringent test of the one-loop ChPT calculations~\cite{loop2} as well as a measure of the energies at which they do not accurately converge.  The data will include both charge channels and a determination of 
the energy dependence of all of the s- and p-wave multipoles for the first time. These experiments are well within the capabilities outlined in Sec.~\ref{sec:facilities}. They should include a precise measurement of the unitary cusp in the region of the $\pi^{+}n$ threshold at 151.4 MeV. They should provide a first measurement of the $\pi$-N phase shifts in the neutral channels. This includes a measurement of the s-wave charge exchange scattering length $a_{cex}( \pi^{+} n \rightarrow \pi^{0} p)$, which will be a measure of isospin symmetry in the $\pi$N system at the few \% level. 

First we show the order of magnitude expected for the observables.  The cross sections have been shown in Fig.~\ref{fig:dsig_Mainz}. It is seen that the $O(p^{4})$ ChPT calculations\cite{loop2} are in excellent agreement with experiment. However the experiments using polarization have just barely begun. These are needed to stringently test the theory. Using transversely polarized photons and unpolarized targets, one can access 
$A(\vec{\gamma})\equiv \Sigma(\theta) = -R_{TT}/R_T$. This is the smallest of the three time-reversal-even asymmetries. A first measurement of this quantity has been performed at Mainz \cite{Schmidt} 
and the results are shown in Fig.~\ref{fig:Sigma_Mainz} and discussed in the previous subsection(~\ref{sec:previous}).
 This observable is important for determining the p-wave amplitudes and is a significant 
test of ChPT calculations \cite{loop,loop-2,loop-3,loop2}. In the future a   measurement of the energy dependence of this quantity can be performed. 
A full MonteCarlo simulation was performed for the anticipated experiments at 
HI$\gamma$S. The predictions of ChPT were used for the $\pi^0 p$ channel, while
those of the DMT model were employed in the case of the $\pi^+ n$ channel. The
experimental setup utilized the Crystal Box assembly
described in Sec.~\ref{sec:facilities} for  $\pi^0$ detection and
the Blowfish neutron detector array, arranged in a 9$\times$9 assembly (with
a center opening) for neutron detection. The beam-on-target intensity
was taken to be 10$^7$ $\gamma$/s, and the HIFROST target was assumed
with a thickness of 3.5 $\times$ 10$^{23}$ protons/cm$^2$. All 
observables were measured at all CM scattering angles. The observables for
the case of the $\pi^0$p channel which were considered were 
$\sigma_T(\theta)$, $\Sigma(\theta)$, $T(\theta$), $E(\theta$), and $F(\theta$).
The same was true for the $\pi^+$n channel, except that T($\theta$), which
is negligibly small in this channel, was omitted.

\begin{figure}
\begin{center}
\includegraphics[width=2.0in]{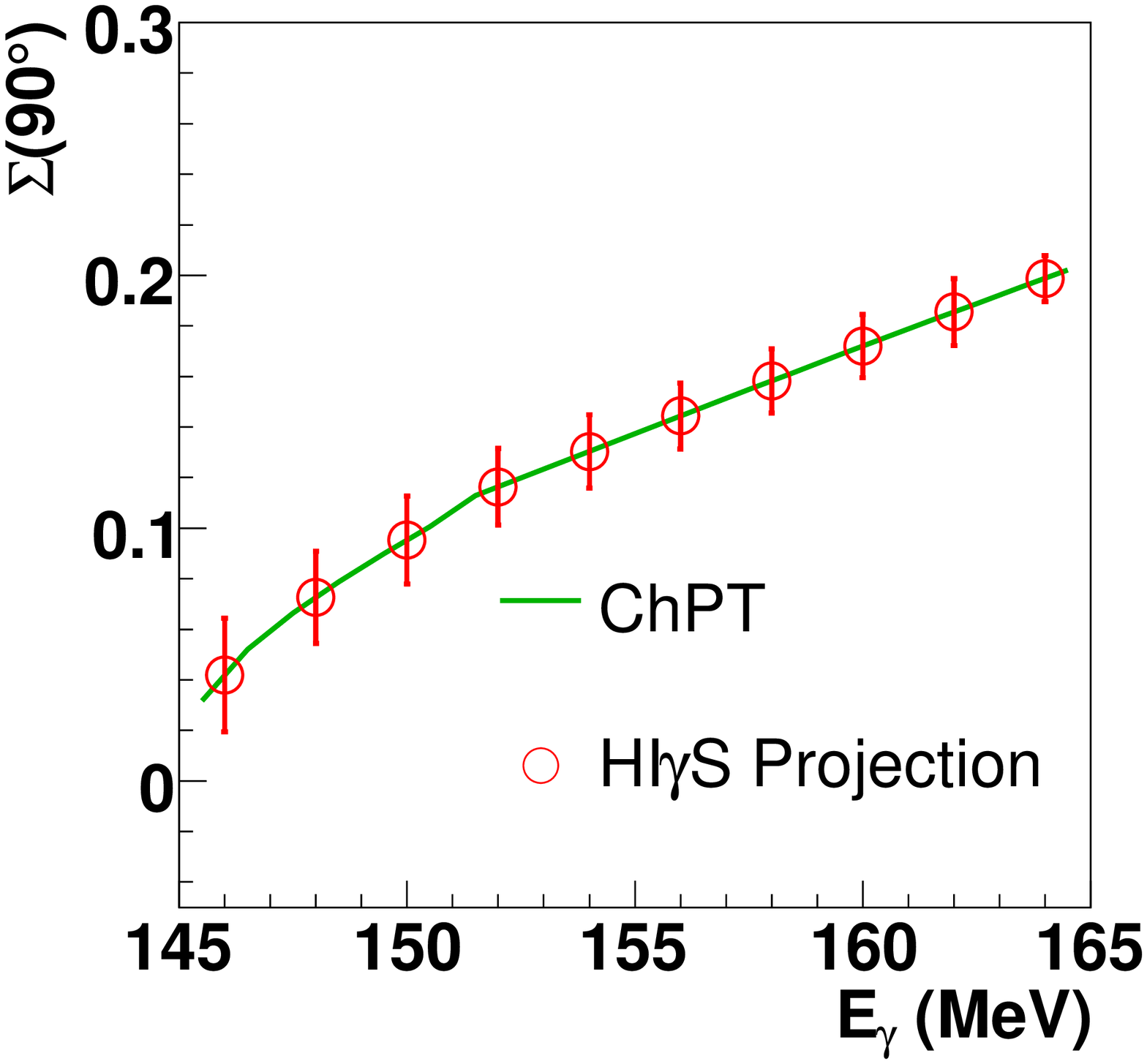}
\includegraphics[width=2.0in]{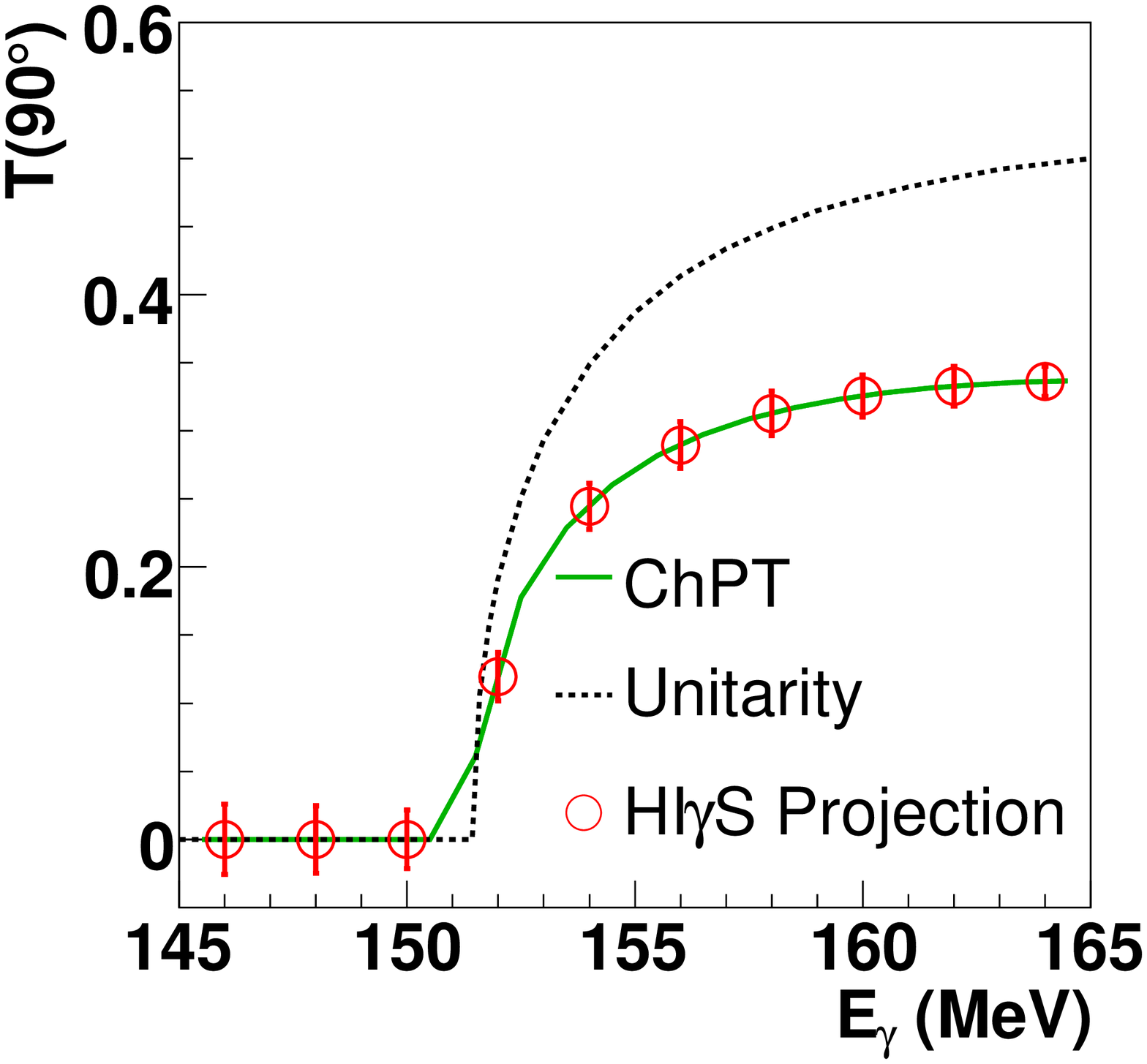}
\includegraphics[width=2.0in]{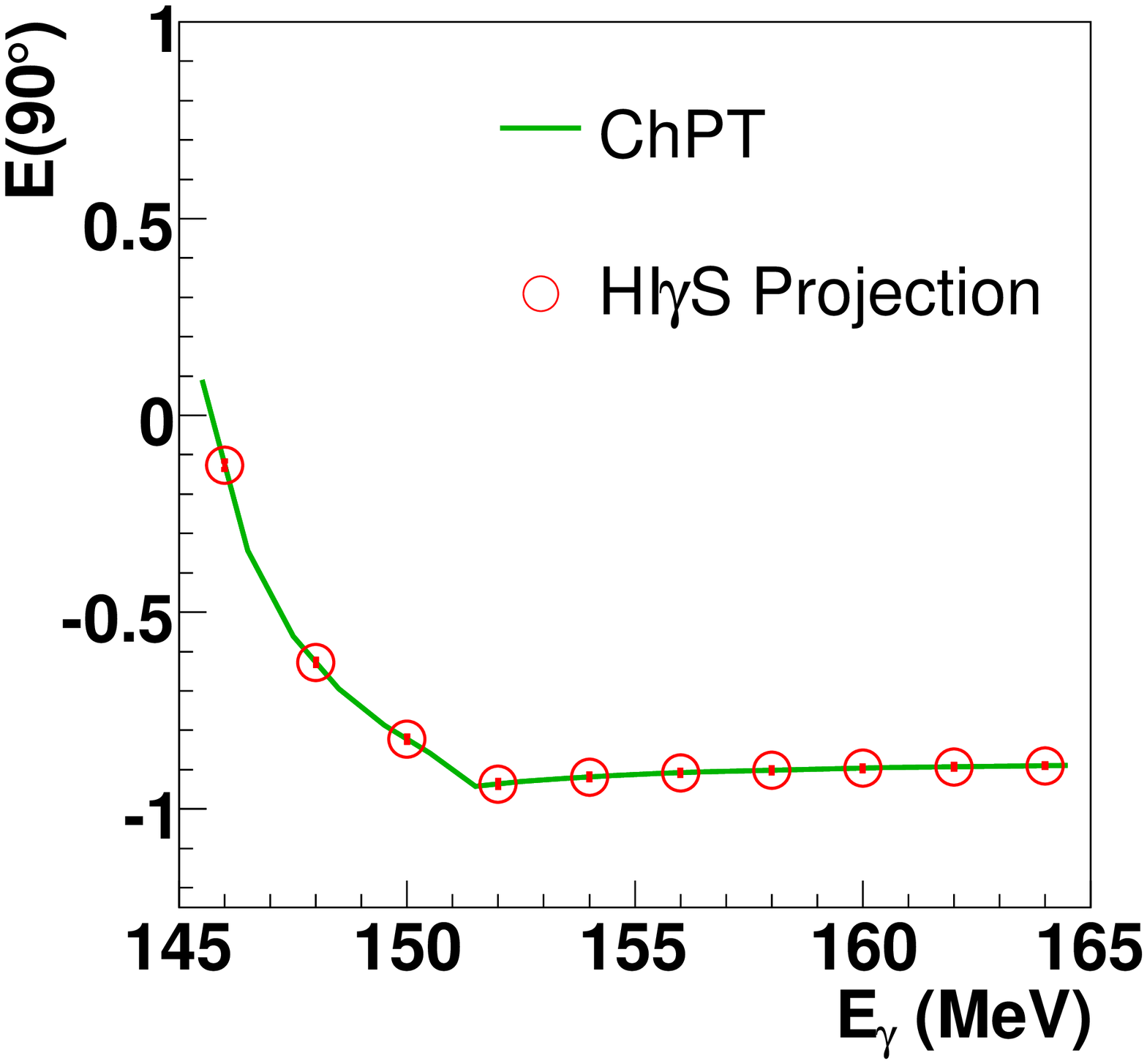}
\includegraphics[width=2.0in]{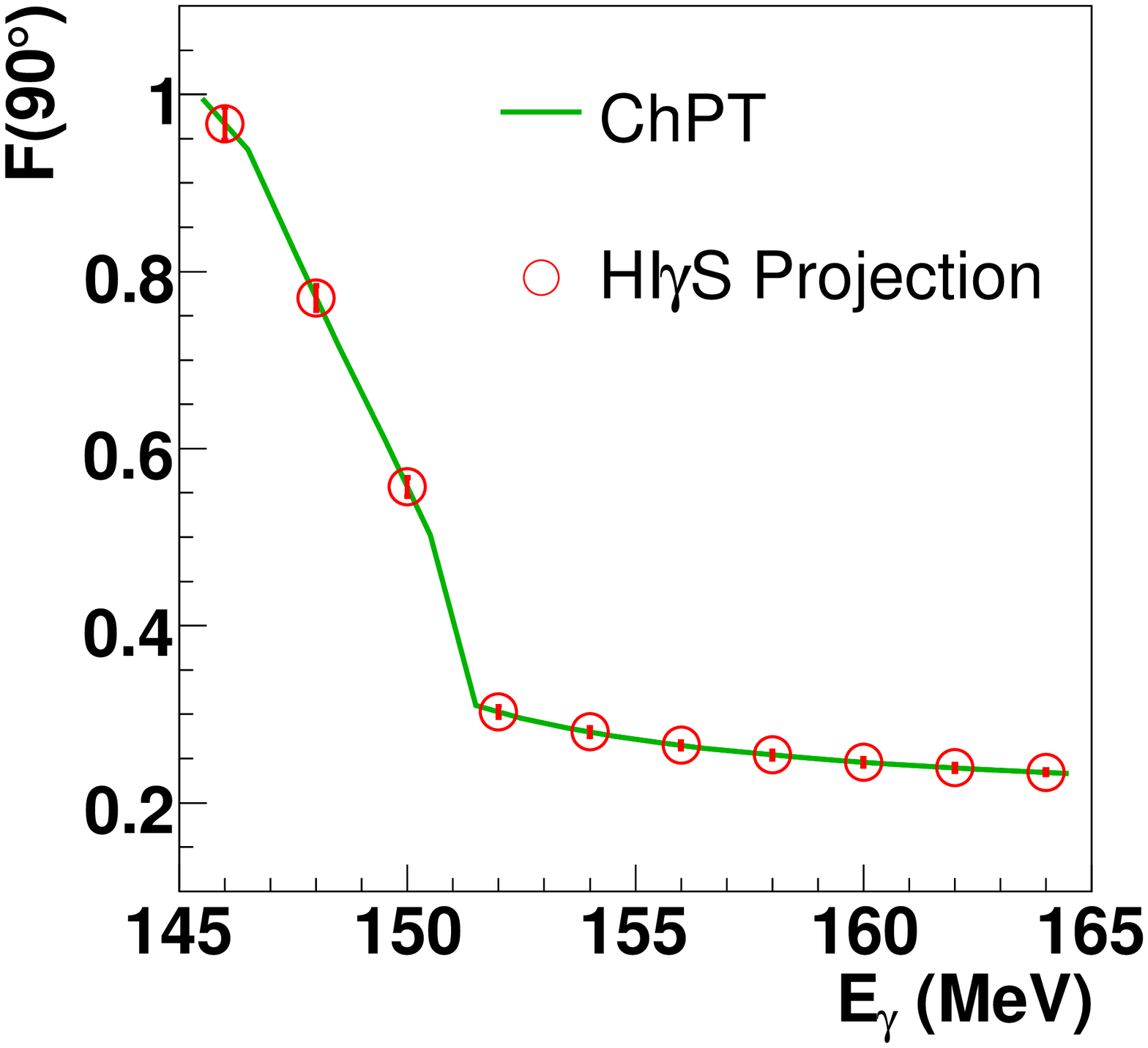}
\end{center}
\caption{ Predicted polarization asymmetries as a function of photon energy at a pion CM angle of 90$^{0}$ for the $\vec{\gamma} \vec{p} \rightarrow \pi^{0}p$ reaction.
The upper panels show the  $\boldsymbol{A(y)\equiv T(90^{\circ})}$ and $A(\gamma)\equiv \Sigma(90^{\circ})$ predictions versus photon energy,
while the lower panels display the predictions for
$E$($90^{\circ}$) and $F$($90^{\circ}$)  versus energy.
 The solid curves starting at the $\pi^{+}n$ threshold correspond to the
 ChPT predictions~\cite{loop2} (see text for discussion).
The dotted curve shown for the case of the $T$($90^{\circ}$) (upper right)
is the result of a fit based on unitarity (see text).
The points show projected statistical errors at various photon energies for 100 hours of running time each at HI$\gamma$S. 
See Table~\ref{table_obs} for definitions.}
\label{fig:pi0-asym}
\end{figure}

The projected observables for the $\pi^0$p channel are shown in Fig.~\ref{fig:pi0-asym}. The estimated single-polarization asymmetries at 90 degrees as a function of photon energy are shown in the upper 
panels of Fig.~\ref{fig:pi0-asym}. Note that $A(\gamma)\equiv \Sigma(90^{\circ})$ 
is non-zero starting at the $\pi^{0}p$ threshold, since it depends on the p-wave amplitudes, which rise smoothly with energy. The plotted points show the estimated errors that will be obtained in runs of $\simeq$ 100 hours each at HI$\gamma$S. Similar errors can be obtained at Mainz, although with longer running times.  

As can be seen in the upper right  panel of Fig.~\ref{fig:pi0-asym} the curves for $\boldsymbol{A(y)\equiv T(90^{\circ})}$ rise rapidly at the $\pi^{+}n$ threshold. This is 
because it is  proportional to Im $ E_{0+}$  which rises rapidly at that threshold due to the unitary cusp. The two 
curves show the large difference in Im $E_{0+}$ for the unitary \cite{AB_lq} ($\beta=3.43$) and ChPT \cite{loop,loop-2,loop-3,loop2} 
calculations ($\beta = 2.78$) discussed in Sec.~\ref{sec:IS_thresh} and shown in Fig.~\ref{fig:E0p}. These curves show the 
sensitivity to $\beta$ that will enable us to measure $a_{cex}(\pi^{+}n \rightarrow \pi^{0}p)$. 

 Predictions for  the two time-reversal-even asymmetries $A(\gamma_{c},z)\equiv E(90^{\circ})$ and 
$A(\gamma_{c},x) \equiv F(90^{\circ})$ are presented in the lower panels of  Fig.~\ref{fig:pi0-asym}. These involve circularly polarized photons and polarized targets. Their values are predicted to be quite large and show the cusp structure in the s-wave production amplitude $E_{0+}$ due to the presence of $\mid E_{0+} \mid^{2}$ and the interference of Re$E_{0+}$ with the p-wave amplitudes. Both of these observables contain a different sensitivity to the s- and p-wave multipoles in comparison with the  unpolarized cross section. 

These will be the first measurements of the imaginary part of $E_{0+}$. The errors that we anticipate for $E_{0+}$ are shown in Fig.~\ref{fig:E0p}. These errors are based upon the assumption that we measure all
four observables shown in Fig.~\ref{fig:pi0-asym} at each energy and all
CM angles. For the real part the significant improvement over what has been previously achieved is evident. Among other issues the  convergence of the ChPT series in photon energy will be tested by these experiments.     

\begin{figure}
\begin{center}
\includegraphics[width=3.0in]{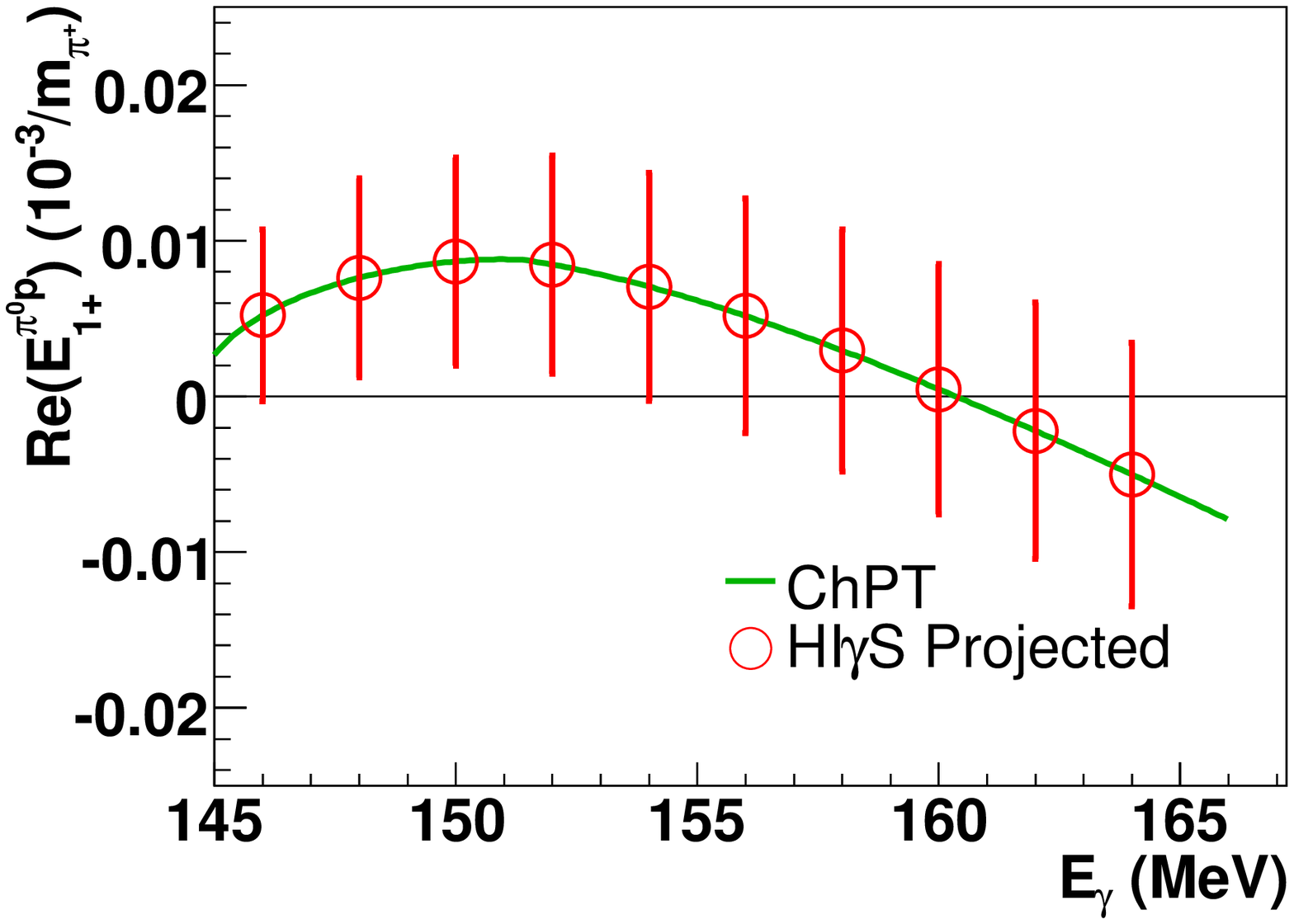}
\includegraphics[width=3.0in]{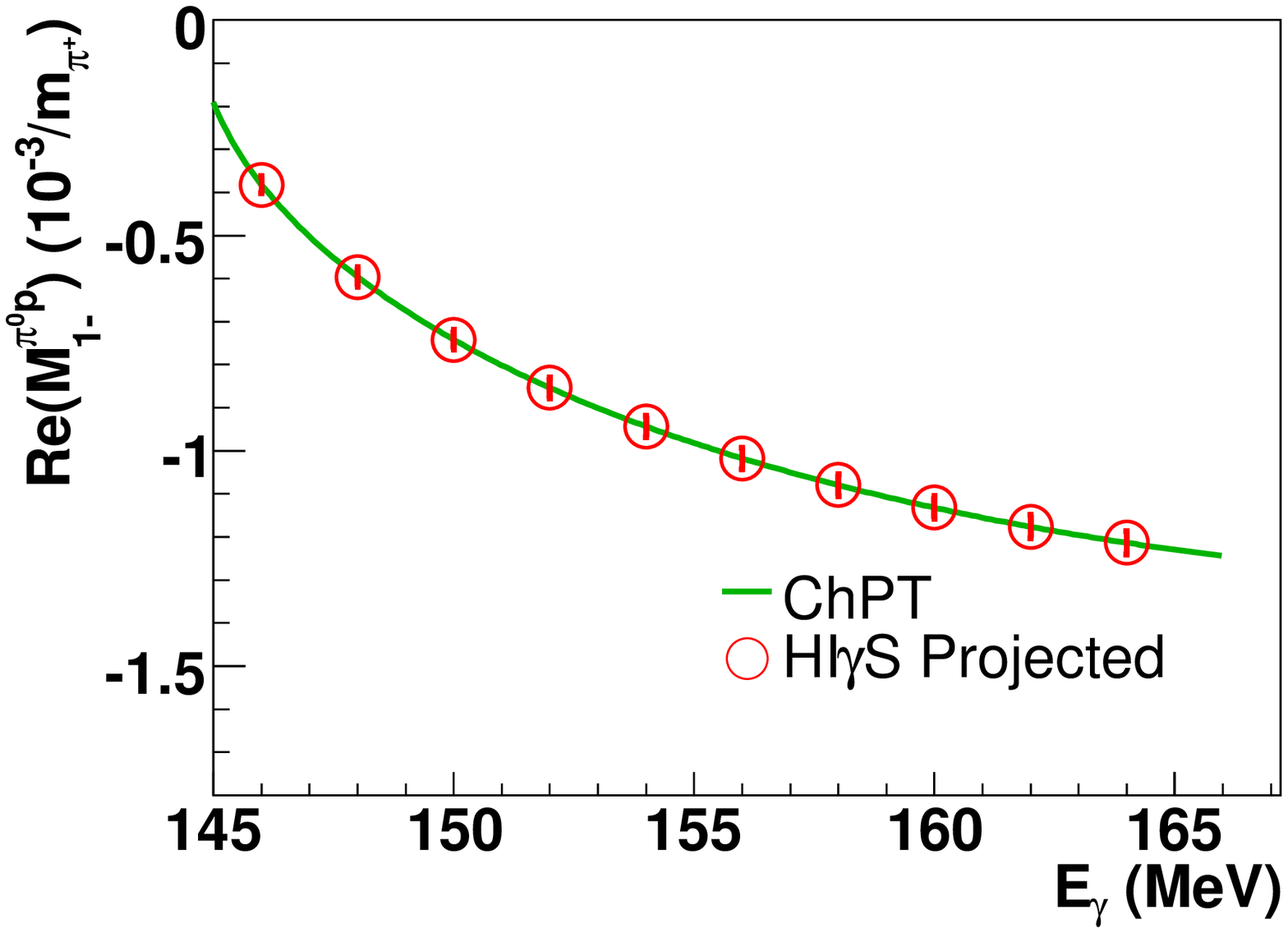}
\includegraphics[width=3.0in]{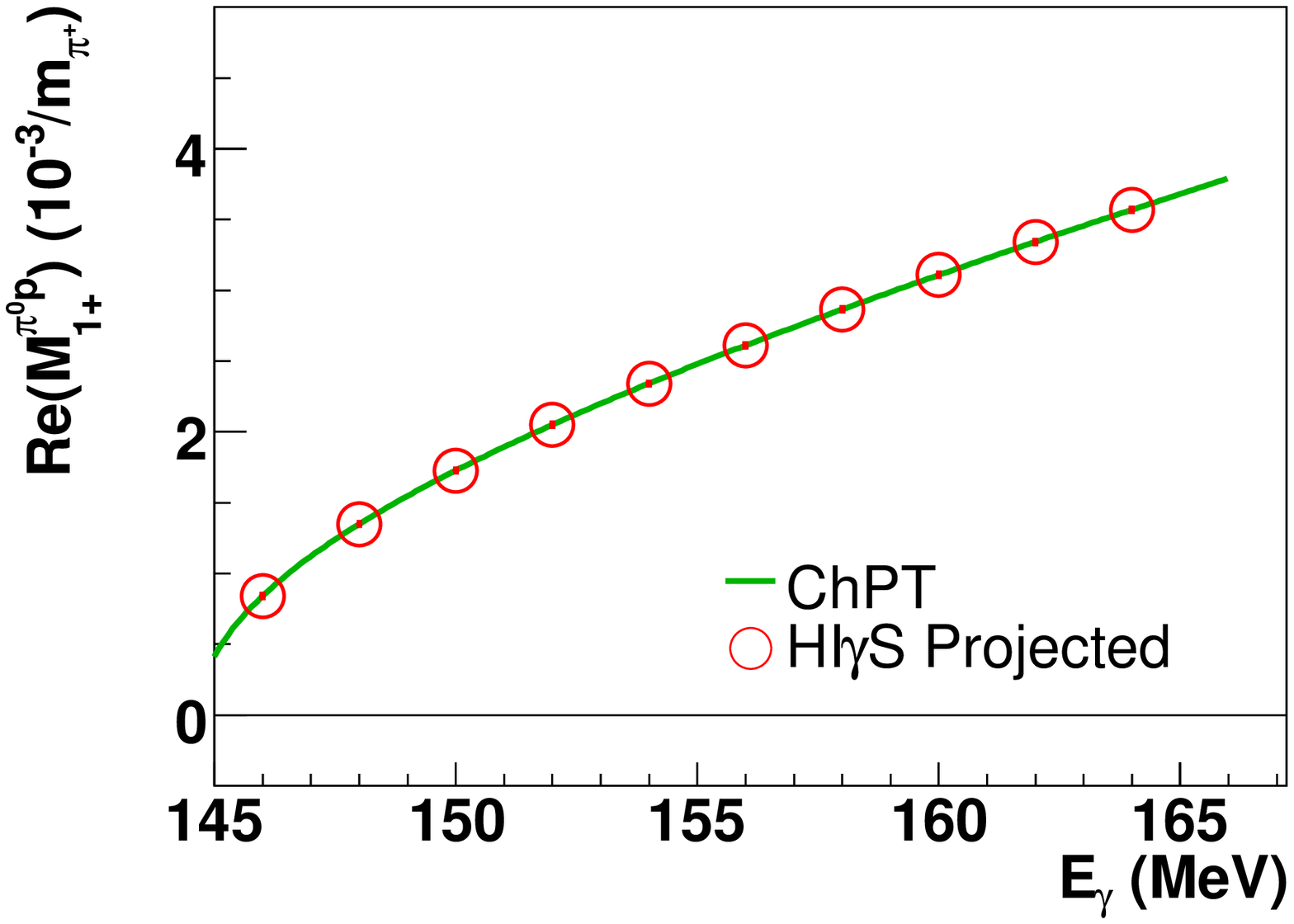}
\end{center}
\caption{Top: Re ($E_{1+}$), Middle: Re ($M_{1-}$), Bottom: Re ($M_{1+}$) for the $\gamma p \rightarrow \pi^{0}p$ reaction. The solid curves are a ChPT \cite{loop,loop-2,loop-3, loop2} calculation. For all panels the projected data points show the estimated statistical errors for 100 hours of beam time at HI$\gamma$S at each energy
in four different beam-target polarization configurations (see Fig.~\ref{fig:pi0-asym}). }
\label{fig:fig8}
\end{figure}

The observables of Fig.~\ref{fig:pi0-asym} also lead to values for the three
p-waves amplitudes. The results are shown in Fig.~\ref{fig:fig8}, along with
to the predictions of ChPT~\cite{loop,loop-2,loop-3,loop2}. Note that although there has been
a previous report for the (small) value of Re $E_{1+}^{\pi^{\circ}p}$
~\cite{PRC55}, but it is highly model dependent since it relies on a
measurement of coherent $\pi^{0}$ photoproduction in Carbon. This will be
the first model independent determination of this small multpole including
is energy dependence and will
provide a very sensitive test of the predictions of ChPT.

Last, but far from least, we discuss one case in which  isospin  breaking is
$\simeq (m_{d} -m_{u})/(m_{d} +m_{u}) \simeq 25\%$ rather then the usual
 $\simeq (m_{d} -m_{u})/ \Lambda_{QCD} \simeq 2\%$. The quantity of interest is
$a(\pi^{0}p)$, the s-wave $\pi^{0}$ scattering length on the proton~\cite{W2,FM}. Clearly this
quantity cannot be directly  measured  since $\pi^{0}$ beams cannot be constructed. Our present
knowledge comes from data involving charged pions and constructing the isospin even s-wave
scattering length $a^{+} =(a_{\pi^{-}p }+ a_{\pi^{+} p})/2= (a_{\pi^{-}p }+ a_{\pi^{-} n})/2$ (it
is understood that the Coulomb contributions have been removed). This can be obtained from
scattering data or more accurately from pionic hydrogen and deuterium~\cite{Meissner-pi-D, Gotta}.
If isospin symmetry holds, then $a^{+} = a(\pi^{0}p)$.
In analyzing the pionic hydrogen and deuterium data it has been shown that the isospin breaking
terms, which are primarily electromagnetic in origin, have to be taken into
account~\cite{Meissner-pi-D}. When this is done, the latest  reported data give the value $a^{+} =
(0.0069 \pm 0.0034)m_{\pi}^{-1}$~\cite{Gotta}.

It has been suggested that $ a(\pi^{0}p)$ can be
measured as a final state interaction in the $\gamma \vec{p} \rightarrow \pi^{0} p$ reaction with
transversely polarized protons in the energy region between the $\pi^{0}$ threshold of 144.7 MeV and
the $\pi^{+}$ threshold of 151.4 MeV~\cite{AB_lq}.
The HI$\gamma$S facility may be capable of a direct measure of this provided that it can produce 
a beam $\simeq 10^{9}$ photons/s on target. A 1000 hour experiment  will enable a measurement of  $
a(\pi^{0}p)$ with a statistical  accuracy $\simeq 10^{-3}/m_{\pi}$, which is comparable with the
present determination of $a^{+}$. A  comparison of $ a(\pi^{0}p)$ and $a^{+}$ will test the
predicted violation of isospin symmetry by $\simeq 25\%$~\cite{W2,FM}. 

\subsection{The $\vec{\gamma} \vec{p}  \rightarrow \pi^{+} n$ Reaction}\label{sec:charged}
For charged pion photoproduction there  is the  low-energy theorem of Kroll and Ruderman~\cite{KR} which leads to an electric dipole amplitude ratio $E_{0+}(\gamma p \rightarrow \pi^{+}n)/ E_{0+}(\gamma p \rightarrow \pi^{0} p) \simeq$ -20. Both the magnitude and sign of this ratio is observable in a unitary cusp which is most visible in the $\pi^{0}$ channel, the strength of which is characterized by the parameter $\beta= 
E_{0+}(\gamma p \rightarrow \pi^{+} n)a_{cex}(\pi^{+} n \rightarrow \pi^{0} p)$
(see Sec.~\ref{sec:IS_thresh}).  

There is only one modern $\gamma p \rightarrow \pi^{+}n$ experiment in the near-threshold region~\cite{Korkmaz}. This 
 was  performed at Saskatoon by neutron detection
yielding the threshold value  $E_{0+}(\gamma p \rightarrow \pi^{+}n)= (28.06 \pm 0.27 \pm 0.45) \times 10^{-3}/m_{\pi^+}$ 
where the first error is statistical and the second is systematic. This is in good agreement with the predictions of the Kroll-Ruderman term plus the chiral corrections
~\cite{KR-chiral} of $28.2\pm0.6$  using the ``old" value of the $\pi$N coupling constant $f_{\pi N}^{2}=0.079 \pm 0.002$ (2.1\%). 
Since the dominant Kroll-Ruderman term is proportional to $f_{\pi N}$, the experimental value of $E_{0+}$ can be used to extract the value of  $f_{\pi N}^{2}=0.078 \pm 0.004$, where the total experimental and theoretical errors are added in quadrature.  This is part way between the old value and the newer value of 
$f_{\pi N}^{2}=0.075 \pm 0.001$~\cite{JJDe}, but the error is too large to distinguish between these two values. In addition we note that the dispersion analysis~\cite{dispersion} gives a value of $E_{0+}(\gamma p \rightarrow \pi^{+} n)$ of 28.0 $\pm$ 0.2 (0.7\% lower) which is significant for this purpose. 
It is clear that an improvement in accuracy for both the theory and experiment are important  to obtain an independent, accurate 
value for the $\pi$N coupling constant (for  further discussion see ~\cite{AB-fpiN}).

\begin{figure}
\begin{center}
\includegraphics[width=2.0in]{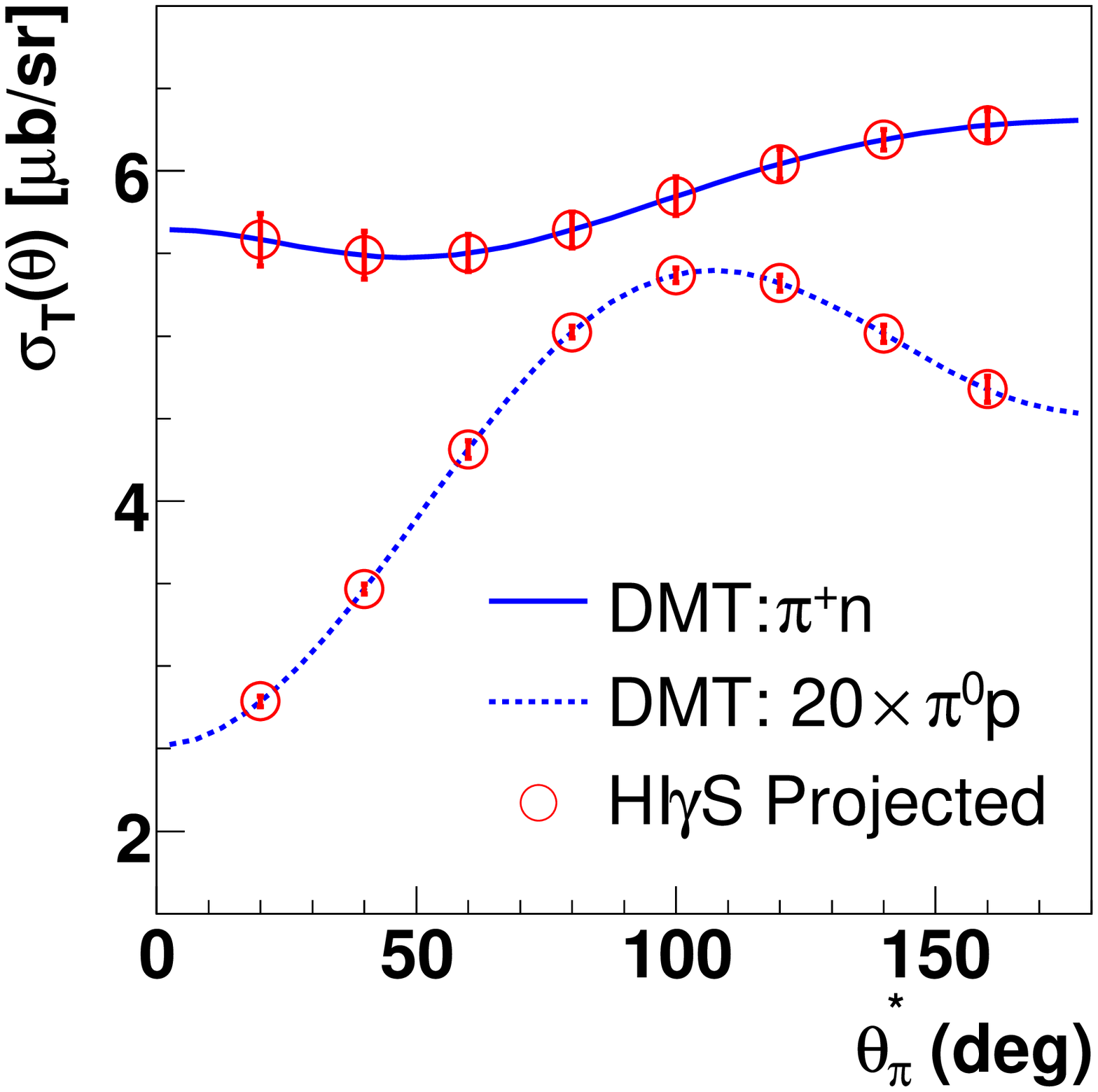}
\includegraphics[width=2.0in]{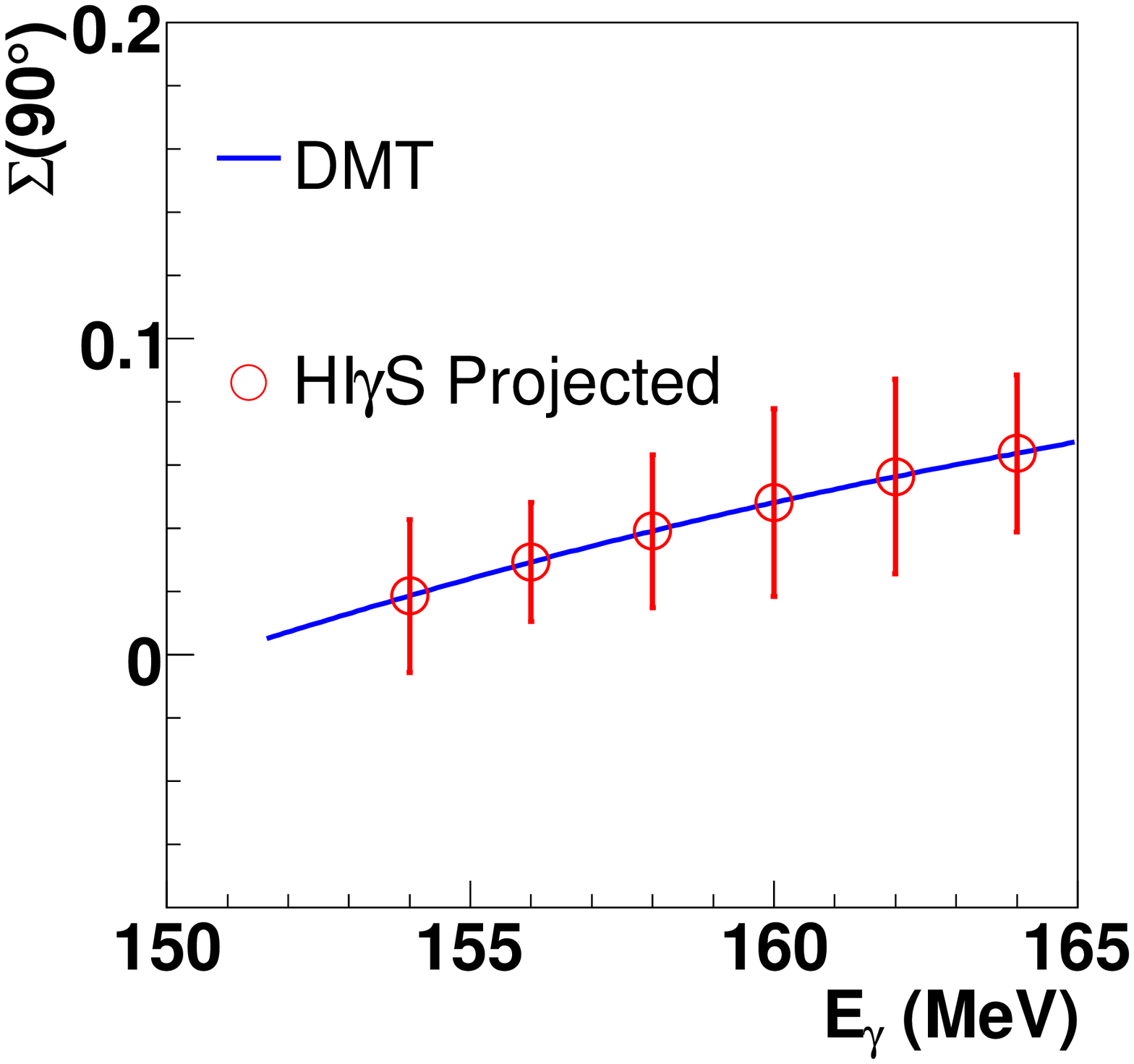}
\includegraphics[width=2.0in]{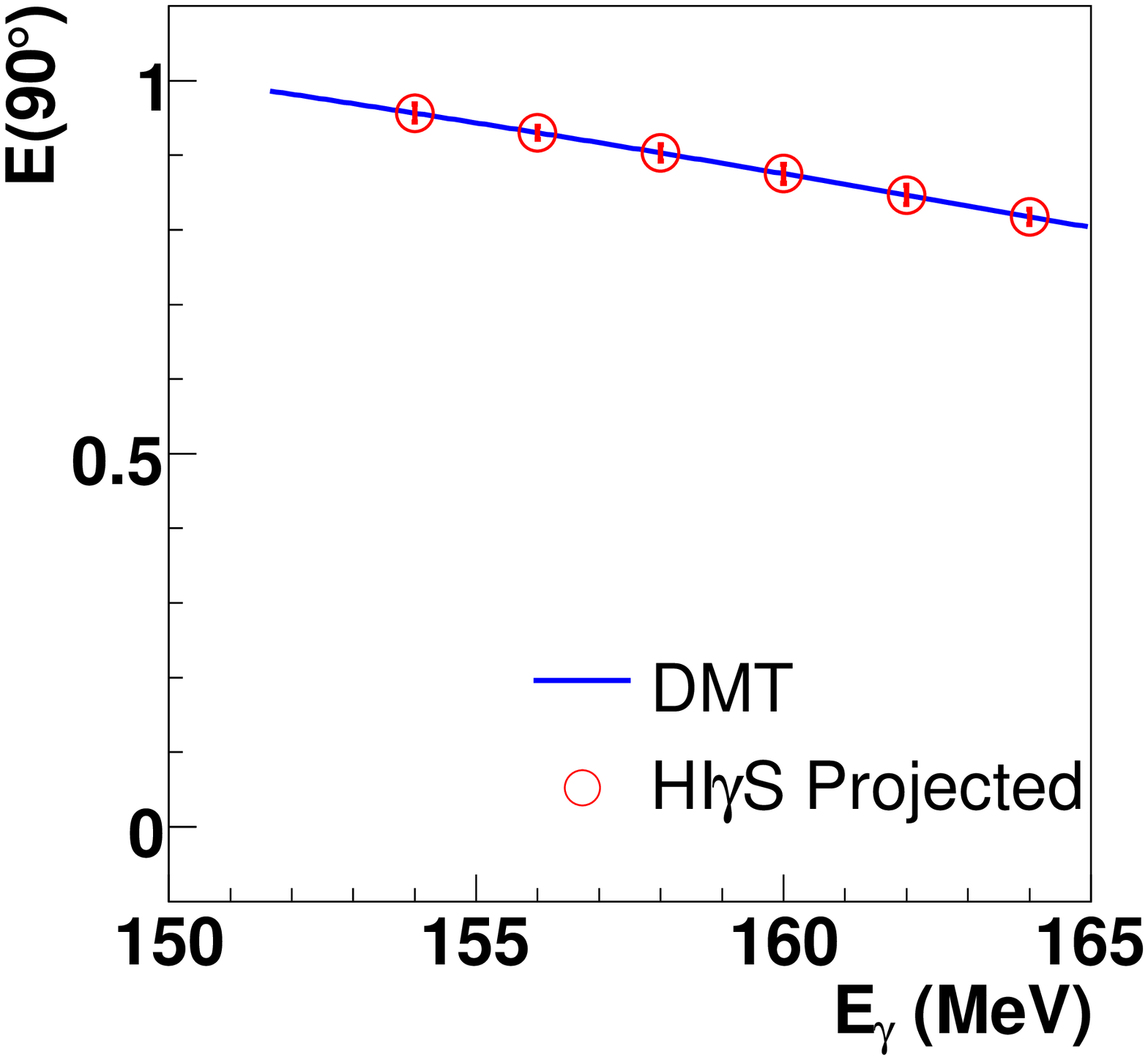}
\includegraphics[width=2.0in]{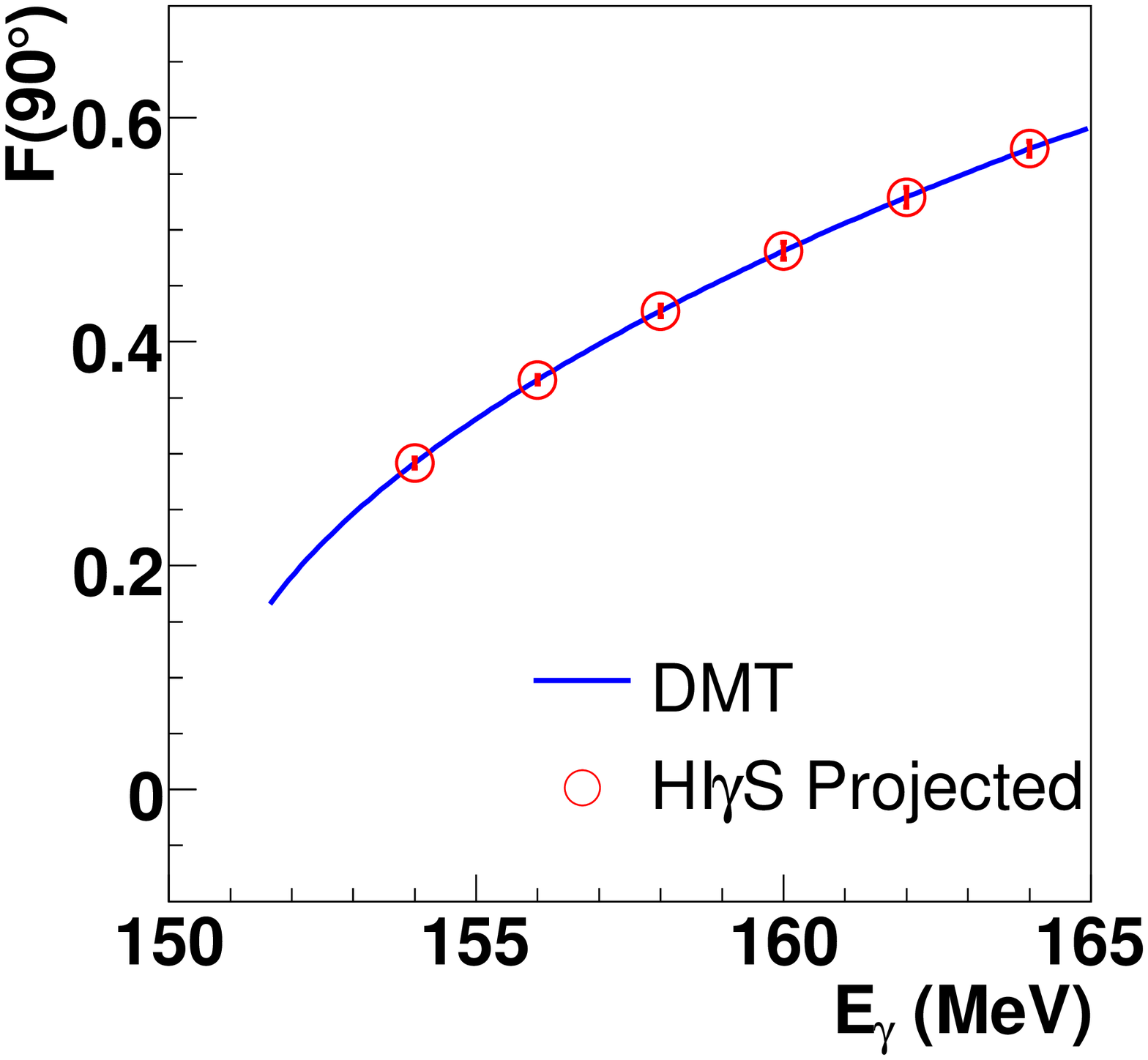}
\end{center}
\caption{Predicted cross section for the $\pi^0 p$ and $\pi^+ n$ channels at a photon energy of 164 MeV (upper left). The time reversal even polarization asymmetries as a function of photon energy at a pion CM 
angle of 90$^{\circ}$ for the $\vec{\gamma} \vec{p} \rightarrow \pi^{+}p$ reaction using the DMT model~\cite{DMT,DMT-2} are shown in the other three panels. 
The data points show the projected statistical uncertainties for 100
hours of running at HI$\gamma$S for each data point.
The calculated time reversal odd asymmetries are all small in this energy range and are not shown. 
See Table~\ref{table_obs} for definitions.}
\label{fig:pip}
\end{figure}

As was discussed above a major motivation to measure $E_{0+}(\gamma p \rightarrow \pi^{+}n)$ accurately is that it is needed to extract the 
value of $a_{cex}(\pi^{+}n \rightarrow \pi^{0}p)$ from the value of $\beta$ measured in neutral pion 
photoproduction~\cite{AB_lq, Anant} (see Sec.~\ref{sec:IS_thresh}). A realistic goal for an accuracy of $\simeq$ 2\% is sufficient for the first round of experiments but it would ultimately need to be improved to reach the $\leq$ 1\% requirement for isospin tests.  In addition, there is considerable interest in exploring 
the energy dependence of $E_{0+}(\gamma p \rightarrow \pi^{+}n)$ in order to  test the chiral 
corrections to the Kroll-Ruderman theorem and which also affect the energy dependence of the $\beta$ parameter in $\pi^{0}$ production.  Furthermore, 
it would be of considerable interest to measure the p-wave multipoles to further test the ChPT 
calculations~\cite{loop2} 
in this channel. Both of these measurements will require full polarization, since both the p-wave 
multipoles and the chiral corrections to the Kroll-Ruderman theorem are relatively small.
The differential cross sections for the $\gamma p \rightarrow \pi^{0}p,\,\pi^{+}n$ 
reactions at a photon energy of 164 MeV are shown in Fig.~\ref{fig:pip}. 
The projected statistical errors in the data points shown are
based on running for 100 hours at HI$\gamma$S for each data point using
the experimental setup at HI$\gamma$S.

Figure~\ref{fig:pip} also displays the results of our MonteCarlo simulation
for the polarization observables $\Sigma(90^{\circ})$, $E(90^{\circ})$, and 
$F(90^{\circ})$ as a function of $\gamma$-ray energy for the $\pi^+ n$ channel.
These projected results are based on the DMT model. Notice that while
$\Sigma(90^{\circ})$ is much smaller here than in the $\pi^0 p$ channel, the
$E(90^{\circ})$ and $F(90^{\circ})$ values are considerable.

\begin{figure}
\begin{center}
\includegraphics[width=5.5in]{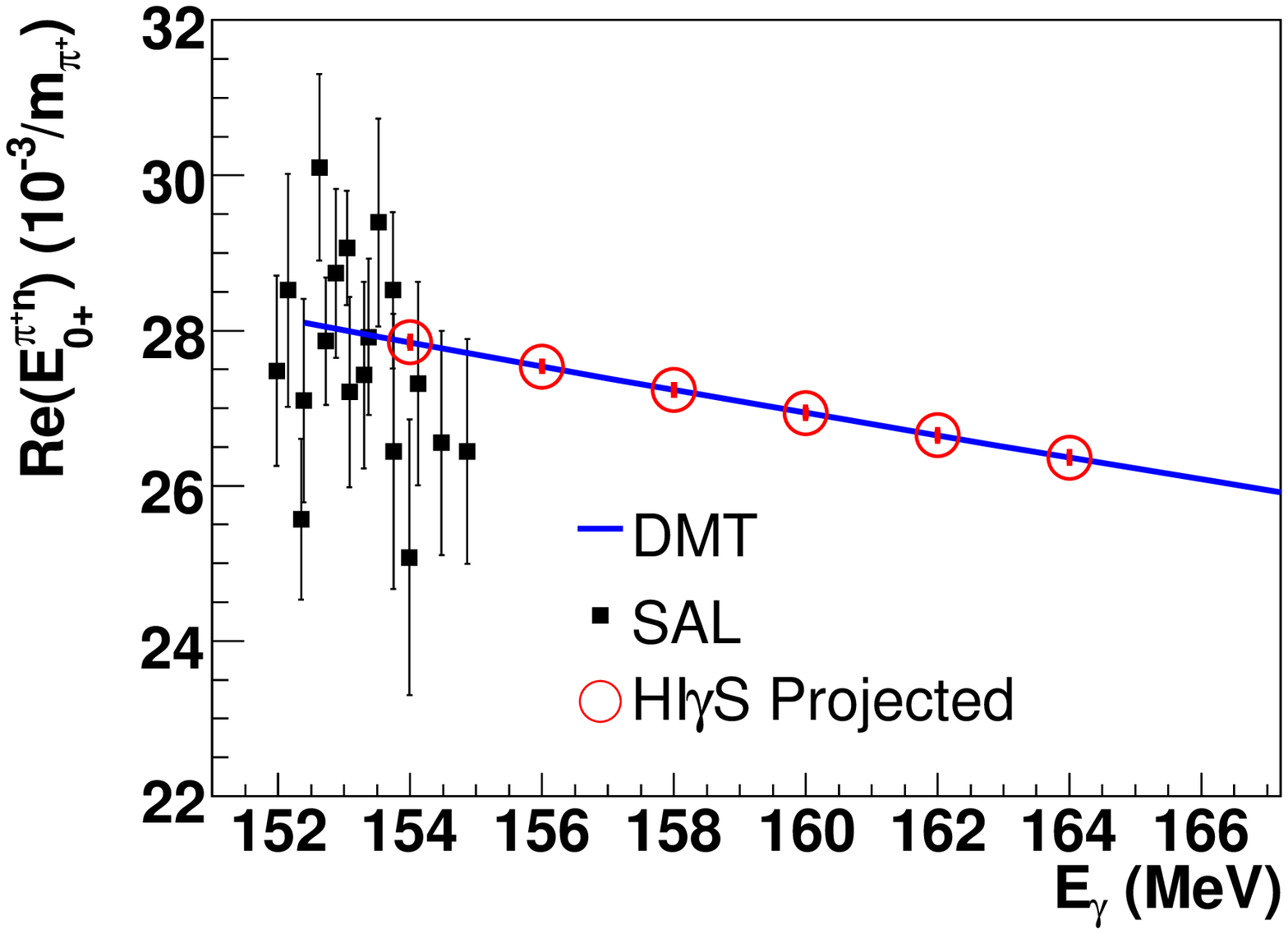}
\end{center}
\caption{Re $(E_{0+})$ for the $\gamma p \rightarrow \pi^+n$ reaction.  The curve is
DMT \cite{DMT} and the square data points are from SAL \cite{Korkmaz}. The
projected data points show the estimated errors for 100 hours of beam time
at HI$\gamma$S at each energy in four different beam-target polarization
configurations (see Fig.~\ref{fig:pip}). }
\label{fig:fig10}
\end{figure}

The results of measuring the four observables shown in Fig.~~\ref{fig:pip}
at essentially all CM angles (there are slight losses at
extreme angles) lead to values for the Re $E_{0+}$ in the
$\pi^+ n$ channel, as shown in Fig.~\ref{fig:fig10}. This result, when combined
with the value of $\beta$ obtained in the $\pi^0$p experiment, will
lead to an accurate value of a$_{cex}$($\pi+ n$ $\rightarrow$$\pi^0 p$)
 (see Eq.~\ref{eq:unitary_cusp}) as well as the value of $f_{\pi n}$, the pion-nucleon
coupling constant.
It can be seen that the charged-pion cross sections are over 20 times larger  than for neutral pion production. This is a consequence of the large Kroll-Ruderman 
term~ \cite{KR,KR-chiral}, which enhances the s-wave amplitude $E_{0+}$ for charged pions. Another consequence of 
this is that $\pi^{0}$ production is p-wave-dominated at these energies. 
This can be seen from the fact  that the differential cross section varies more rapidly with angle for the neutral pion production than for the $\pi^{+}n$ channel.

\subsection{Photo-Pion Production From the Neutron: Charge Symmetry Tests}\label{sec:D}
 To properly study photo-pion production from the nucleon we also need to consider the 
 $\vec{\gamma} \vec{n} \rightarrow \pi^{0}n, \pi^{-}p$ reactions. There are four charge channels, and assuming that isospin is conserved for the strong interactions (i.e. for the final $\pi$N states), there are three independent  isospin amplitudes for each multipole~\cite{obs}. Therefore to completely determine these multipoles experimentally it takes a complete experiment on three of the charge channels. To test isospin conservation it will take a similar experiment on the fourth charge channel. In order to do  this in a significant way  one would need to determine the multipoles at the 1\% level. 

The most straightforward approach to measuring the four reaction channels is to
consider the deuteron as a target. 
This introduces  few body physics which must  be shown to be under control in order to extract accurate neutron target data. On the other hand it provides additional important physics such as  testing effective field theory calculations for the two body system. An important first step in the theory has been taken by the $O(p^{4})$ ChPT calculation of the coherent $\gamma D \rightarrow \pi^{0} D$ reaction at threshold  and its demonstration that the neutron-$\pi^{0}$ contribution is significant~\cite{Beane}. 
 The use of the deuteron target also allows  the study of  the charge symmetry of the  NN interaction at low energies~\cite{D-ann} which is also fundamental since any charge asymmetry is due in part to the up, down quark mass difference.  
 
One  channel which involves the neutron, but not the deuteron, is the ``inverse" $\pi^{-} p \rightarrow \gamma n$ reaction for which there is some data from TRIUMF between threshold and the $\Delta$ resonance~\cite{TRIUMF-inverse,TRIUMF-inverse-2}. On the theoretical side there
 is  an  $O(p^{3}$) ChPT calculation~\cite{Fearing} which is in reasonable agreement with all of the  data~\cite{TRIUMF-inverse}. Furthermore it should be noted that if the assumption of resonance saturation is employed in ChPT calculations, the low energy parameters derived from the proton data can be used to predict the two charge channels on the neutron~\cite{loop}. Clearly the photopion amplitudes from the neutron are as fundamental as those for the proton, just more difficult to access experimentally. This is a future challenge for the field. 
 
 Few body calculations for the deuteron can be tested for many reactions such as $\gamma D \rightarrow \pi^{0}D, \pi^{0} np, \pi^{+}nn, \pi^{-} pp$. In addition there are special kinematic regions where the two outgoing nucleons move together with a small relative momentum. It is  this regime that is very sensitive to the $a_{NN}$: the three nucleon-nucleon scattering lengths. There is a long history of measurements and calculations on this subject (for a review see~\cite{charge-symm}). It has been concluded that there is strong evidence for a charge symmetry violation because $a_{nn}- a_{pp} $ is not equal to zero~\cite{charge-symm}. However there has been a long controversy over the value of $a_{nn}$. Many of the reactions used to measure this quantity have three interacting hadrons in the final state. The radiative capture reaction $\pi^{-} p \rightarrow n n \gamma$ does not have this problem and an accurate value of $a_{nn}$ has been extracted using it~\cite{pion-ann}. It has been shown by accurate modern calculations that the $\gamma D \rightarrow n n \pi^{+}$ reaction should also be capable of obtaining an accurate value~\cite{D-ann}. This reaction has the advantage that there are certain ranges of the kinematic parameters that are insensitive to $a_{nn}$ and which therefore can be used to test the reaction  calculations. In addition the two other reactions $\gamma D \rightarrow \pi^{0} np, \pi^{-}pp$ can be used to measure $a_{np}$ and $a_{pp}$ which have previously been accurately measured~\cite{charge-symm}. This will provide a strict test of this method. Accurate data on the three N-N scattering lengths are another way to get at the isospin breaking dynamics due to the mass difference of the up and down quarks~\cite{V_NN-symmetry}.

\section{Conclusions}\label{sec:conclusions}  
We have described a significant program of photo-pion physics that can be 
carried out at the HI$\gamma$S facility and at Mainz.  Specifically, we can achieve the following:
\begin{itemize}
\item{A first full measurement of the unitary cusp in the real and imaginary parts of the s-wave electric dipole amplitude $E_{0+}(\gamma p \rightarrow \pi^{0}p)$  from a 
measurement of the transverse polarized target asymmetry $\boldsymbol{A(y) = T(\theta)}$. This will provide the energy dependence of Im$E_{0+}$ and  an accurate measurement of 
$\beta = E_{0+}(\gamma p \rightarrow \pi^{+}n) a_{cex}(\pi^{+}n \rightarrow \pi^{0}p)$. With our  value of 
$E_{0+}(\gamma p \rightarrow \pi^{+}n)$, we will obtain a first measurement of 
$a_{cex}(\pi^{+}n \rightarrow \pi^{0}p)$ at the few \% level 
of accuracy. As has been discussed in Sec.~\ref{sec:IS_thresh}, this will 
be a test of isospin breaking, which has been predicted due to the mass difference of the 
up and down quarks $m_{d}-m_{u}$.}
\item{The first  measurement of the energy dependence of $A(\gamma)\equiv \Sigma(\theta)$ for the  
$\gamma p \rightarrow \pi^{0}p$ reaction. Combined with  accurate cross-section 
measurements~\cite{Schmidt} this will enable us to obtain the 
energy dependence of the three p-wave multipoles, which will test the theoretical calculations \cite{loop2,DMT,DMT-2}.}
\item{ We have shown that there is a realistic possibility to measure
$a(\pi^{0}p)$, the s-wave $\pi^{0}$ scattering length on the proton, as a
final state interaction in the $\gamma \vec{p} \rightarrow \pi^{0} p$
reaction with transversely polarized protons in the energy region between
the $\pi^{0}$ threshold of 144.7MeV and the $\pi^{+}$ threshold of 151.4
MeV\cite{AB_lq} to a state-of-the-art accuracy of $\simeq 10^{-3}/m_{\pi}$.
This quantity is predicted to have an isospin breaking  $\simeq (m_{d}
-m_{u})/(m_{d} +m_{u}) \simeq 25\%$ rather then the usual
 $\simeq (m_{d} -m_{u})/ \Lambda_{QCD} \simeq 2\%$\cite{ W2,FM}. }
\item{A more accurate measurement of the cross section for the $\gamma p \rightarrow \pi^{+}n$ reaction 
just above threshold. From this, we can obtain a more  accurate value (1-to-2 \%) for  $E_{0+}(\gamma p \rightarrow \pi^{+}n)$ 
and the $\pi$N coupling constant (f$^2_{\pi N}$) at the same level of
accuracy, as well as $a_{cex}(\pi^{+}n \rightarrow \pi^{0}p)$, as discussed above.}
\item{A first measurement of the asymmetries in the $\gamma p \rightarrow \pi^{+}n$ reaction just above 
threshold. Combined with the accurate measurement of $E_{0+}(\gamma p \rightarrow \pi^{+}n)$, this will 
enable us to obtain the p-wave multipoles in this channel for the first time. This can be used as a further 
test of the theoretical calculations~\cite{loop2,DMT,DMT-2}.}
\item{ The measurement of $\pi$ N scattering lengths, and the possible
checking of a large isospin breaking at intermediate energies, indicates the
sensitivity of photo pion reactions to the pion-nucleon final state when
transversely polarized targets are employed. This shows the ability to
measure $\pi$ N phase shifts in states with a minimum electromagnetic
interaction.}

\end{itemize} 

These experiments for the $\vec{\gamma}\vec{p} \rightarrow \pi^{0}p,\, \pi^{+}n$ reactions with photon energies 
just above the $\pi^{+}n$ threshold (151.4 MeV) up to $\simeq$ 160 MeV are planned for near-term operation of 
the HI$\gamma$S facility and at Mainz. The intermediate-energy isospin-breaking experiments discussed in Sec.~\ref{sec:IS} can be performed 
at Mainz and, as soon as the upgrades outlined in this paper have been completed, at the
HI$\gamma$S facility. We believe that these data will provide stringent new tests of ChPT, and probe the isospin breaking due to the mass difference of the up and down quarks. 

\section{Acknowledgments}
The work described in this paper is supported in part by US Department of Energy, Office of Science grants DE-FG02-97ER41033 and DE-FG02-94ER40818, and US Department of Defense MFEL Program contract number FA9550-04-01-0086. We would also like to thank
V. Bernard, U.G. Meissner, and C. Fernandez-Ramirez for their comments on 
the manuscript and discussions.

\begin{appendix}
\section{s- and p-Wave Multipole Expansion of the Response Functions}
For convenience we have collected the formulas for some of the response
functions in terms of the s and p wave multipole amplitudes starting with the time reversal even (RE) observables: The $A_{T},
B_{T}, C_{T}$ coefficients of $\sigma_T$ and the polarized photon(unpolarized target) responses are:
\begin{equation}
\label{eq:ABC}
\begin{array}{c}
        R_T(\theta) = A_T + B_T \cos\theta +C_T \cos^2 \theta \\
        A_{T}= {\mid}E_{0+}{\mid}^{2} + {\mid}P^{+}_{23}{\mid}^{2} \\
        B_{T}= 2 Re( E_{0+} P_{1}^{\star}) \\
        C_{T}= \mid P_{1}{\mid}^{2} -{\mid}P^{+}_{23}{\mid}^{2}\\
         R_{TT} = \sin^{2} \theta {\mid}P^{-}_{23}{\mid}^{2} 
\end{array}
\end{equation}
Here we ave introduced the ChPT notation for the p wave
multipoles:
\begin{equation}
\label{eq:P}
\begin{array}{c}
        P_{1}= 3  E_{1+} + M_{1+} - M_{1-} \\
        P_{2}= 3  E_{1+} - M_{1+} + M_{1-} \\
        P_{3}= 2 M_{1+} + M_{1-} \\
{\mid}P^{\pm}_{23}{\mid}^{2}= ({\mid}P_{2}{\mid}^{2} \pm {\mid} P_{3}{\mid}^{2})/2
\end{array}
\end{equation}

\begin{equation}
\label{eq:circ}
\begin{array}{c}
R_{TT'}^{0z}   =   a_{TT'}^{0z} + b_{TT'}^{0z}\cos\theta + c_{TT'}^{0z}\cos^2\theta\\
a_{TT'}^{0z} = -|E_{0+}|^2 - Re(P_2^* P_3)\\
b_{TT'}^{0z} = -2 Re(E_{0+}^* P_1)\\
c_{TT'}^{0z} = Re(P_3^* P_2) - |P_1|^2
\end{array}
\end{equation}

\begin{equation}
\begin{array}{c}
       R_{TT^{'}}^{0x} = \sin \theta Re [ (E_{0+}^{*} + \cos\theta  P_{1}^{*})(P_{2}-P_{3})] 
\end{array}
\end{equation}

For the largest time reversal odd observable
\begin{equation}
\begin{array}{c}
    R_{T}^{0y} = \sin \theta Im[ ( (E_{0+}^{*} + \cos\theta  P_{1}^{*})(P_{2}-P_{3})] 
 \end{array}
\end{equation}
Note that this is the imaginary amplitude of $R_{TT^{'}}^{0x}$.

These all can be derived from the complete formulas using the CGLN F invariant amplitudes expanded into s and p waves\cite{MAID-eta}.
\begin{equation}
\begin{array}{c}
 F_{1} = E_{0+} +3(M_{1+} +E_{1+}) \cos\theta = 
 E_{0+} +(P_{1} +P_{3}) \cos\theta \\
 F_{2} = M_{1-} +2 M_{1+} = P_{3}\\
 F_{3} = 3(E_{1+} -M_{1+}) = P_{2} -P_{3} \\
 F_{4} = 0
 \end{array}
\end{equation}
\end{appendix}


\end{document}